\documentclass[12pt]{article}
\usepackage{graphicx}
\usepackage{latexsym}
\usepackage{epsfig}

\newcommand{\abs}[1]{\left | #1 \right |}
\newcommand{\beq}{\begin{equation}}
\newcommand{\eeq}{\end{equation}}
\newcommand{\beqa}{\begin{eqnarray}}
\newcommand{\eeqa}{\end{eqnarray}}
\newcommand{\lsim}{\begin{array}{c}\,\sim\vspace{-21pt}\\< \end{array}}
\newcommand{\gsim}{\begin{array}{c}\sim\vspace{-21pt}\\> \end{array}}

\newcommand{\capt}[1]{%
     \begin{minipage}[t]{5in}%
     \caption{#1}%
     \end{minipage}%
     }

\newcommand{\LL}{L}
\newcommand{\GF}{\Lambda}

\newcommand{\Dslash}[1]{\!\not{\!#1}\,}
\newcommand{\sla}[1]{\!\!\not{\!#1}}

\begin{document}

\begin{titlepage}

\begin{center}


\begin{flushright}
\normalsize{
UCB-PTH-03-01\\
LBNL-52029\\
FERMILAB-PUB-03/017-T
  }
\end{flushright}

\vspace{0.5cm}

{\Large \bf Bulk Gauge Fields in Warped Space and \\
Localized Supersymmetry Breaking}

\bigskip

{\bf Z. Chacko}$^{\bf a,b,c}$,
and {\bf Eduardo Pont\'{o}n}$^{\bf d,e}$ \\

\medskip

$^{\bf a}${\small \it Department of Physics, University of California,
Berkeley, CA 94720, USA \\
\medskip
$^{\bf b}$ Theoretical Physics Group, Lawrence Berkeley National Laboratory, \\
Berkeley, CA 94720, USA \\
\medskip
$^{\bf d}$ Fermi National Accelerator Laboratory,
P.O. Box 500, Batavia, IL 60510, USA \\

\medskip

$^{\bf c}${\rm email}: zchacko@thsrv.lbl.gov $\; \; \;$ $^{\bf e}${\rm email}: 
eponton@fnal.gov} \\

\vspace{0.5cm}

{\bf Abstract}

\end{center}
\noindent

We consider five dimensional supersymmetric warped scenarios in which 
the Standard Model quark and lepton fields are localized on the 
ultraviolet brane, while the Standard Model gauge fields propagate in 
the bulk.  Supersymmetry is assumed to be broken on the infrared 
brane.  The relative sizes of supersymmetry breaking effects are found 
to depend on the hierarchy between the infrared scale and the weak 
scale.  If the infrared scale is much larger than the weak scale the 
leading supersymmetry breaking effect on the visible brane is given by 
gaugino mediation.  The gaugino masses at the weak scale are 
proportional to the square of the corresponding gauge coupling, while 
the dominant contribution to the scalar masses arises from 
logarithmically enhanced radiative effects involving the gaugino mass 
that are cutoff at the infrared scale.  While the LSP is the 
gravitino, the NLSP which is the stau is stable on collider time 
scales.  If however the infrared scale is close to the weak scale then 
the effects of hard supersymmetry breaking operators on the scalar 
masses can become comparable to those from gaugino mediation.  These 
operators alter the relative strengths of the couplings of gauge 
bosons and gauginos to matter, and give loop contributions to the 
scalar masses that are also cutoff at the infrared scale.  The gaugino 
masses, while exhibiting a more complicated dependence on the 
corresponding gauge coupling, remain hierarchical and become 
proportional to the corresponding gauge coupling in the limit of 
strong supersymmetry breaking.  The scalar masses are finite and a 
loop factor smaller than the gaugino masses.  
The LSP remains the gravitino.

\end{titlepage}

\renewcommand{\thepage}{\arabic{page}}
\setcounter{page}{1}

\section{Introduction}

Supersymmetry provides an attractive framework for solving the 
hierarchy problem but it introduces several naturalness puzzles of its 
own.  In particular it is not clear why the squark masses should be 
flavor diagonal.  One class of natural solutions arises in theories 
where the quark and lepton fields are localized on a `3-brane' in 
extra dimensions and the hidden sector field which breaks 
supersymmetry is localized on another spatially separated 3-brane.  In 
such a framework contact terms between the visible and hidden fields 
in the four dimensional effective theory generated by the exchange of 
bulk states with masses much larger than the inverse brane spacing are 
exponentially small \cite{Randall:1998uk}.  The dominant contribution 
to the soft scalar masses will then be generated by the exchange of 
the light states in the bulk.  If the couplings of these light states 
are flavor blind the scalar masses will be flavor diagonal.  The 
spectrum of the superparticles obtained for various choices of the 
bulk states has been extensively investigated in the literature for 
the case when the space between the branes is flat (unwarped).  When 
the higher dimensional supergravity multiplet is the only light bulk 
state then the leading contribution to the soft scalar masses is given 
by anomaly mediation \cite{Randall:1998uk,Giudice:1998xp}.  If the 
MSSM gauge fields are also in the bulk and couple directly to the 
supersymmetry breaking sector then the leading contribution to the 
soft scalar masses arises from gaugino mediation 
\cite{Kaplan:1999ac,Chacko:1999mi} provided the hidden sector field 
which breaks supersymmetry is a singlet.  If this field is not a 
singlet then other effects such as radion mediation 
\cite{Chacko:2000fn} and gaugino assisted anomaly mediation 
\cite{Kaplan:2000jz} dominate.  In each of these cases a 
characteristic and interesting spectrum of flavor diagonal soft masses 
is obtained.{\footnote{It may not be straightforward to realize these 
setups in the context of string/M theory {\cite 
{Anisimov:2001zz}}{\cite {Anisimov:2002az}}.}}

In this paper we consider contributions to the superparticle masses 
when the space between the branes is not flat, but warped as in RS1 
{\cite{RS1}}.  We have two major motivations for considering 
supersymmetry breaking effects in these spaces.  Firstly, since any 
realistic brane is expected to have a tension some degree of warping 
seems inevitable, and since the effect of warping on the spectrum can 
be significant it should be taken into account.  Secondly, as we 
establish in this paper, physics on any one of the branes is screened 
from symmetry breaking effects on the other brane above the 
compactification scale, which is of order the mass of the lightest 
Kaluza-Klein particle.  The potentially large hierarchy between the 
cutoff of the theory and the compactification scale (which is in 
contrast to theories with a flat extra dimension) can then be used to 
address other naturalness problems of the Standard Model.

In our framework the Standard Model quarks and leptons are localized 
to the brane where the warp factor is large (the `UV brane') while 
supersymmetry is broken on the brane where the warp factor is small 
(the `IR brane').  The Standard Model gauge fields are in the bulk 
\cite{Goldberger:1999wh,hewettpomarol} and couple directly to the 
supersymmetry breaking sector, while the field which breaks 
supersymmetry is a singlet.  We further assume that the radion is 
stabilized supersymmetrically so that its F component is zero.  Gauge 
coupling constant evolution in these models has been established as 
logarithmic \cite{Pomarol:2000hp}--\cite{Agashe:2002pr}.  Therefore, 
in order to explain the observed unification of coupling constants, we 
further restrict ourselves to the case where the theory respects a 
grand unifying symmetry at very short distances that is broken by the 
Higgs mechanism on the UV brane at a scale of about $10^{16}$ GeV.

In general, we find that the relative sizes of various supersymmetry 
breaking effects depend on the warp factor and therefore the spectrum 
differs significantly from that of the corresponding theory in flat 
space.  If the warp factor is sufficiently small that the 
compactification scale is hierarchically larger than the weak scale 
then the dominant contribution to the Standard Model superpartner 
masses arises from gaugino mediation.  Here the gaugino masses arise 
from a nonrenormalizable contact operator on the IR brane and are 
proportional to the square of the corresponding gauge coupling.  The 
scalar masses are finite and generated radiatively through loops 
involving the gaugino mass.  These loops are cutoff at the 
compactification scale so that while the scalar masses are loop 
suppressed with respect to the gaugino masses they are logarithmically 
enhanced by the ratio of the compactification scale to the weak scale.  
We determine the spectra for several different values of the 
compactification scale.  In these models the LSP is the gravitino, 
while the NLSP is the stau.  Since the scale of supersymmetry breaking 
is relatively high, the stau is stable on collider time scales.  Some 
degree of fine tuning is required to obtain a viable spectrum, in part 
because of the bound on the Higgs mass and in part because of the 
bound on the stau mass.

If the warp factor is sufficiently large that the compactification 
scale is close to the weak scale then contributions to the scalar 
masses from loop diagrams involving hard supersymmetry breaking 
operators can become large and comparable to the effects of the 
gaugino mediation.  The effect of these hard breaking operators, which 
were previously considered in \cite{Kaplan:2000jz}, is to alter the 
relative coupling of gauge bosons and gauginos to matter.  While the 
new loop diagrams contributing to the scalar masses are finite and 
cutoff at the compactification scale like the gaugino mediated graphs, 
their contributions are no longer related to the gaugino masses, and 
further, their sign is not fixed.  The dependence of the gaugino 
masses on the corresponding gauge coupling is also more complicated 
since the effect of repulsion of the gaugino wave function from the IR 
brane is in general no longer negligible.  The gaugino masses remain 
hierarchical and in the limit of strong supersymmetry breaking become 
proportional to the corresponding gauge coupling.  Supersymmetry 
breaking in this limit where the compactification scale is close to 
the weak scale has been considered previously 
\cite{GP1,GP2,Marti:2001iw,Goldberger:2002pc}.  However these authors 
did not take into account the effects of brane localized kinetic terms 
on the soft masses.  In models where the grand unifying symmetry is 
broken on the UV brane, either by the Higgs mechanism or by boundary 
conditions, these effects cannot be neglected since the difference in 
the four dimensional gauge couplings of SU(3), SU(2) and U(1) arises 
from such terms.  As a result our conclusions differ from theirs in 
significant details.

This paper is organized as follows.  In section 2 we describe the 
general properties of fields propagating in an AdS background.  We
also address some aspects of the quantum theory and the physics of 
localized gauge symmetry breaking.  In section 3 we turn our attention 
to localized supersymmetry (SUSY) breaking and determine the gaugino 
masses in various limits.  In section 4 we discuss the phenomenology 
of the present class of models and in section 5 we give our 
conclusions.  We summarize various technical details in the 
appendices.  In appendix A we derive the gauge boson propagator in the 
presence of localized gauge symmetry breaking, and we present its 
behavior in various energy regimes.  In appendix B we derive the 
gaugino propagator including the breaking of the gauge symmetry.  
Finally, in appendix C we derive the lightest gaugino mass when SUSY 
breaking is localized on the infrared brane and there are brane 
localized (gauge boson and) gaugino kinetic terms on the UV brane.

\section{The Framework}

We now consider a concrete five dimensional scenario.  We employ a 
coordinate system $x^M$ where $M$ runs from 0 to 3 and 5.  The fifth 
dimension $x^5 = y$ is compactified on the interval $0 \le y \le
\LL$, which can be thought as arising from the orbifold $S^1/Z_2$.  
There are 3-branes at the orbifold fixed points $y =0$ and $y = \LL$.  
The space is warped and the Standard Model quarks and leptons are 
localized on the brane at $y =0$ where the warp factor is large.  
Supersymmetry is assumed to be broken on the brane where the warp 
factor is small.  The metric is given by the line element
\begin{equation}
\label{lineelement}
ds^2 = e^{-2 \sigma} \eta_{\mu \nu} dx^{\mu}dx^{\nu} + dy^2~.
\end{equation}
Here the $x^{\mu}$, where $\mu$ runs from 0 to 3, parametrize our usual
four spacetime dimensions, $\eta_{\mu \nu} = {\rm{diag}}(-1,1,1,1)$,
and $\sigma = k \abs{y}$, where $k$ is the AdS curvature and is related
to the four dimensional Planck scale $M_{4}$ and the five dimensional
Planck scale $M_{5}$ by
\beq
\label{4DPlanckscale}
M_4^2 = \frac{M_5^3}{2k}\left(1 - e^{-2 k \LL} \right) \simeq 
\frac{M_5^3}{2k}~,
\eeq
where the second equality holds when the warping is significant.

The gauge fields of the Standard Model are assumed to propagate in the 
bulk.  An on shell vector multiplet in five dimensional ${\cal{N}} = 
1$ supergravity consists of a gauge field $A_M$, a pair of symplectic 
Majorana spinors $\lambda^i$, with $i=1,2$, and a real scalar $\Sigma$ 
which transforms in the adjoint representation.  The bosonic part of 
the higher dimensional gauge field Lagrangian takes the 
form\footnote{The minus sign in front of the scalar kinetic term is 
due to our metric signature convention, Eq.~(\ref{lineelement}).}
\beq
{\cal{L}}_{b} = -\frac{1}{g_5^2} \left[\frac{1}{4} F_{MN} F^{MN} +
\frac{1}{2} D_M \Sigma \; D^M \Sigma + \left( \sigma'' -2 k^2
\right) \Sigma^2 \right]~,
\eeq
where $\sigma'' = 2k \left[\delta(y)-\delta(y-\LL) \right]$. The 
fermionic part takes the form
\beq
\label{gauginoaction}
{\cal{L}}_{f} = - \frac{i}{2g_{5}^{2}} \left[\bar{\lambda}^i \Gamma^M D_M
\lambda^i + \frac{1}{2} \sigma'
\bar{\lambda}^i \left( \sigma^{3} \right)^{ij} \lambda^j \right]~,
\eeq
where $D_{M}$ is a covariant derivative with respect to both general 
coordinate and gauge transformations.  The vielbein factors necessary 
to write the spinor action in curved space are implicit in 
Eq.~(\ref{gauginoaction}).  In addition to the interactions above, 
additional brane localized terms that preserve $N = 1$ supersymmetry 
are also possible, but we postpone their consideration until the next 
subsection.

We demand that $A_{\mu}$ and $\lambda^1_L$ are even while $\Sigma$ and 
$\lambda^2_L$ are odd.  Then the even fields each have a massless mode 
with the following $y$ dependence:
\beqa
A_{\mu}\left(x,y\right) &=& \frac{1}{\sqrt{\LL}} A_{\mu}^{\left(0\right)}
\left(x\right) + \cdots \\
\label{gauginozeromode}
\lambda^1_L\left(x,y\right) &=& \frac{e^{3\sigma/2}}  {\sqrt{\LL}}
\lambda^{1 \; \left(0 \right)}_L\left(x\right) + \cdots
\eeqa

In the previous theory, an approximate expression for the masses $m_n$ 
of the Kaluza Klein (KK) states can be obtained in the limit that $k 
\gg m_n$ and $k\LL \gg 1$:
\beq
m_n \approx \left( n - \frac{1}{4} \right) \pi k e^{-k \LL}~.
\eeq
This shows that the KK masses are of order $M_c = k\,e^{-k \LL}$, which
we will call the compactification scale.

In order to make physical predictions for low energy observables we 
will need a consistent framework for quantum field theory in warped 
space.  It has been shown by Goldberger and Rothstein 
\cite{Goldberger:2002cz} that a setup in which the parameters of the 
Lagrangian are defined in terms of correlators with external points 
localized on the Planck brane is one such consistent framework.  In 
particular this allows the renormalization group evolution of the 
theory from short distances down to the compactification scale, where 
it can be matched onto a four dimensional effective theory.  This is 
the approach we shall be using.  Its consistency relies on the the 
general properties of the bulk tree-level propagators.  In fact, all 
perturbative calculations can be done, in principle, once the higher 
dimensional propagators are known.  Even though, in practice, it is 
technically simpler to do calculations in terms of KK modes, an 
understanding of how the higher dimensional calculation would proceed 
provides considerable physical insight.

\subsection{General Properties of Propagators in AdS backgrounds}

To illustrate the physical properties of propagators in the AdS 
background Eq.~(\ref{lineelement}), we consider the tree-level gauge 
boson propagator.  The action has the following form:\footnote{We 
included explicit factors of two in front of the $\delta$-function 
terms for later convenience.  Since we are taking the point of view 
that the physical fifth dimension runs from $y=0$ to $y=\LL$, the 
$\delta$-functions, located at the endpoints of this interval, 
contribute a factor of $1/2$ when performing the $y$ integration.}
\beqa
\label{action1}
&& \hspace{-9mm}  \int_0^\LL d^4\!x \,dy \left\{- \frac{1}{4g_5^2} 
\sqrt{-G}\,F_{MN} F^{MN} + 2\delta\left(y\right) \sqrt{-G_{\rm{ind}}}
\left[ {\cal{L}}_{UV} - \frac{1}{4g_{UV}^2} F_{\mu \nu} F^{\mu \nu} \right]
\right. \nonumber \\
&& \left. \hspace{4.1cm} \mbox{}
+ 2\delta\left(y - \LL \right) \sqrt{-G_{\rm{ind}}}  \left[ {\cal{L}}_{IR} -
\frac{1}{4g_{IR}^2} F_{\mu \nu} F^{\mu \nu} \right] \right\}~,
\eeqa
where ${\cal{L}}_{UV}$ and ${\cal{L}}_{IR}$ describe brane localized 
fields that may be charged under the gauge group.  We also include 
brane localized gauge kinetic terms, which are expected to be induced 
radiatively.  For simplicity, in this section we will assume that they 
can be treated perturbatively (see \cite{Carena:2002me, Carena:2002dz} 
for a discussion of their effects when they are not small.)

When calculating the propagator, it is convenient to work in 
four-dimensional momentum space, as defined by
\beq
\label{Fourier}
G(y;p) = \int \! d^{4}x \, e^{i \eta_{\mu\nu} p^{\mu} x^{\nu}} G(x,y)~.
\eeq 

Keeping the $y$ dependence makes the locality properties along the 
fifth dimension explicit.  The four-dimensional momentum $p$, as 
defined in Eq.~(\ref{Fourier}), is conserved in physical processes in 
the background Eq.~(\ref{lineelement}).  Nevertheless, it is important 
to keep in mind that $p^{\mu}$ is not necessarily the physical 
momentum of the propagating field as measured by an observer standing 
at an arbitrary point in the bulk $y$.  For example, such an observer 
would measure an energy (frequency) $E_{\rm{phys}} = 1/\Delta \tau = 
e^{k y}/\Delta x^{0} = e^{k y} p^{0}$, where $\tau$ is proper time and 
$x^{0}$ is the time coordinate used in Eq.~(\ref{lineelement}), which 
is conjugate to $p^{0}$.  In the following, any statements regarding 
energy scales will refer to the values of the ``coordinate" momentum 
$p$, and therefore they should be understood as scales measured by UV 
observers.

We start with the propagator in the absence of gauge symmetry breaking 
or brane localized gauge kinetic terms, which has been calculated 
elsewhere (see, e.g.  \cite{GP2}).  It is useful to consider first the 
case where the UV and IR boundaries are sent to infinity, so that the 
background is pure AdS.{\footnote{The boundary conditions are then 
defined by the procedure of removing the branes to infinity.}} 
Neglecting the tensor structure and working in Euclidean space, the 
propagator evaluated at coincident points $y = y'$ exhibits a 
remarkably simple behavior (see appendix A)
\beq
\label{AdSpropagator}
G_p(y,y) \sim - \frac{g_5^2}{2} \left \{
\begin{array}{ll}
\frac{e^{k y}}{p} & \vspace{2mm}
\hspace{1cm} {\rm{for}} \;\; p \gg k \, e^{-k y} \\
\frac{2 k}{p^2 \ln(2k/p)} & \hspace{1cm} {\rm{for}} \;\; p \ll k \, e^{-k y}
\end{array}
\right.
\eeq
We note that for sufficiently low energies, the propagator exhibits a 
four-dimensional scaling $\sim 1/p^{2}$, up to a logarithm.  We stress 
that this is the case even before compactification.  This property is 
intimately related to the AdS/CFT conjecture, which states that the 
five-dimensional theory is equivalent to a four-dimensional conformal 
field theory (CFT).  The logarithm that appears for energies below $k 
\, e^{-k y}$ is sometimes described as a tree-level ``running" of the 
gauge coupling and corresponds, in the four-dimensional dual picture, 
to CFT loop effects.  We also see that $G_p(y,y)$ changes from a 
four-dimensional ($\sim 1/p^{2}$) to a five-dimensional ($\sim 1/p$) 
behavior at a scale $p \sim k \, e^{-k y}$.  The onset of the higher 
dimensional scaling indicates that, for localized observers at $y$, 
the theory ceases to be predictive when the external 4-d momenta are 
much larger than $k \, e^{-k y}$.  As emphasized in 
\cite{Goldberger:2002cz}, the breakdown of the theory for external $p 
\gg k\,e^{-k y}$ is already apparent when considering tree-level 
insertions of higher dimension operators localized at $y$, which give 
contributions proportional to powers of $p/(k e^{-ky})$.  Therefore, 
the theory defined by Eq.~(\ref{action1}) should be understood as an 
effective theory with a $y$-dependent cutoff $\Lambda \, e^{-k y}$ on 
$p$, for some constant $\Lambda \gsim k$.{\footnote{Note that this is 
equivalent to a $y$-independent cuttoff on $p_{phys} = p \, e^{k y}$ 
given by $\Lambda$.}}

Now we bring in the UV and IR boundaries.  The asymptotic behavior 
given in Eq.~(\ref{AdSpropagator}) is modified in the following 
manner.  The UV boundary affects only the high energy behavior of the 
propagator when evaluated at (or very close to) $y=0$: in this case 
the propagator is twice the result given in Eq.~(\ref{AdSpropagator}).  
The IR brane, on the other hand, changes only the low energy regime, 
for all $y$.  One finds that for $p \ll k \, e^{-k \LL}$ the $\ln(2k/p)$ 
factor should be replaced by $k \LL$. Thus, the tree-level ``running''
stops at about the compactification scale $k\,e^{-k \LL}$.

One may wonder if a theory with a $y$-dependent cutoff can be used to 
make predictions at all, since even if one is only interested in 
calculating observables at a fixed position $y$, say on one of the 
orbifold fixed points, at some energy scale $p \ll k \, e^{-k y}$, 
such observables can receive contributions from other points in the 
bulk, where the local cutoff is lower than the relevant external 
momentum $p$.  Such contributions are simply not calculable within the 
framework of the five-dimensional theory.  To illustrate this, 
consider the contribution to UV brane observables from the following 
tower of operators localized on the IR brane:
\beq
\label{higherdimops}
{\cal{L}}_{IR} \supset \sum_n c_n F_{\mu\nu} \Box^n F^{\mu\nu}~,
\eeq
where $\Box = G_{\rm{ind}}^{\mu\nu} \partial_\mu \partial_\nu = 
e^{2k\LL} \eta^{\mu\nu} \partial_\mu \partial_\nu$.  The $n=0$ term 
corresponds to the IR brane localized gauge kinetic term that we wrote 
in Eq.~(\ref{action1}), but in general we expect all the operators in 
Eq.~(\ref{higherdimops}) (and many other) to be present in the 
effective theory.  Writing the coefficients of these higher dimension 
operators as $c_n = \hat{c}_n/M_5^{2n}$, where the $\hat{c}_n$ are 
dimensionless, we see that treating these operators perturbatively 
gives a contribution proportional to $\sum_n \hat{c}_n 
[p/(M_5\,e^{-k\LL})]^{2n}$.  This shows that for $p \gsim M_5\,e^{-k 
\LL}$ the effective theory breaks down.  In spite of this observation, 
one should not necessarily conclude that the UV observables are not 
calculable when $M_5\,e^{-k L} \ll p \ll M_5$.  One may see this by 
considering the tree-level propagator for non-coincident points $y\neq 
y'$.  Assuming that $y<y'$, one finds
\beqa
G_p(y,y') &\sim& - \frac{g_5^2 k}{p^2 \ln(2k/p)} \, \sqrt{\frac{\pi p}{2 k}}
\, e^{\frac{1}{2}k y'} e^{-\frac{p}{k} e^{k y'}}
\hspace{8mm} {\rm{for}} \;\;  k \, e^{-k y'} \ll p \ll k \, e^{-k y}~,
\nonumber
\eeqa
which exhibits a huge exponential suppression $e^{-\frac{p}{k} e^{k 
y'}}$.  This suggests that, whatever the physics at $y'$ for energies 
much larger than the local cutoff $M_5\,e^{-k y'}$ is, its effect on 
the physics at $y$ is exponentially suppressed and therefore 
unimportant.  In particular, we do not expect contributions from short 
distance physics at $y'$ to overcome the exponentially small 
probability for a gauge field to propagate there and back.

\subsection{Quantum Field Theories in a slice of AdS}

We now explain how quantum field theories can be formulated in warped 
spaces in terms of propagators with external legs on the Planck brane.  
Our discussion closely follows that of Goldberger and Rothstein 
{\cite{Goldberger:2002cz}}.  For simplicity the theory we will be 
considering is a (non-supersymmetric) bulk U(1) gauge theory with 
fermionic matter localized on the UV and IR branes.  This will suffice 
to illustrate our framework.  The Lagrangian takes the form of 
Eq.~(\ref{action1}) with ${\cal{L}_{UV}}$ and ${\cal{L}_{IR}}$ given 
by
\beq
{\cal{L}}_I =  \bar{\psi}_{I}
\left( i \Gamma^{\mu}\partial_{\mu} - m_{I} \right){\psi}_{I}
\hspace{1cm} I = UV, IR
\eeq
Our regularization scheme involves dimensional regularization with 
momentum subtraction.  Although this scheme is mass dependent it has 
the advantage that the decoupling of heavy states occurs 
automatically, without the necessity to match and run 
{\cite{Manohar:1996cq}}.  Since the cutoff for physics on the UV brane 
is of order $M_5$, the parameters of the Lagrangian ${\cal{L}}_{UV}$ 
on the UV brane can be defined at any subtraction point $\mu$ less 
than this.  This is also true for the parameters of the bulk 
Lagrangian.  However since the cutoff on the IR brane is of order 
$M_5\,e^{-k\LL}$ the parameters of the Lagrangian ${\cal{L}}_{IR}$ on 
the IR brane cannot be defined at any subtraction point higher than 
this.  Nevertheless correlators with external legs on the UV brane and 
four momenta much larger than the cutoff on the IR brane can still be 
calculated up to exponentially small corrections.  This is because, as 
we saw in the previous subsection, a propagator from the UV brane to 
any point in the bulk where the four momentum is larger than the local 
cutoff dies away exponentially quickly.

It is possible to renormalization group evolve the parameters of the 
Lagrangian on the UV brane ${\cal{L}}_{UV}$ and the bulk Lagrangian 
from $\mu$ of order $M_5$ down to the compactification scale 
$k\,e^{-k\LL}$.  Similarly the parameters of the Lagrangian on the IR 
brane can be evolved from $\mu$ of order $M_5\,e^{-k\LL}$ down to the 
compactification scale.  The complete higher dimensional Lagrangian 
can then be matched on to the four dimensional effective theory.  
Below we illustrate this procedure for the U(1) theory.

In order to obtain the dependence of the gauge coupling constant on
the renormalization scale, we consider the quantum corrections to the 
two-point correlator for gauge fields with external points localized
on the UV brane.  For a U(1) gauge theory the corrections to the gauge
coupling constant from renormalization of the vertex function and the
two-point functions of the fermions can be made to cancel by the
Ward-Takahashi identity.  The corrections to the gauge field two point
correlator arise from fermionic loops of the form below which are 
familiar from four dimensions.  Considering first the correction from
the fermionic loop on the UV brane alone we find
\beq
G_{\mu \nu}\left(0,0;p\right) \rightarrow
G_{\mu \nu}\left(0,0;p\right) +
G_{\mu \alpha}\left(0,0;p\right)\left[p^2 \eta^{\alpha \beta} - p^{\alpha} p^{\beta}
\right] i \Pi(p^2) G_{\beta \nu}\left(0,0;p\right)~,
\eeq
where the gauge boson propagator $G_{\mu\nu}$ is related to the 
$G_{p}$ of the previous section by Eq.~(\ref{Gmunu}) of appendix A, 
and $\Pi(p^2)$ must be regulated.  We regulate this by momentum 
subtraction at a scale $p^2 = \mu^2$, which is spacelike due to the 
signature implicit in the line element Eq.~(\ref{lineelement}).  Then 
$\Pi(p^2) \rightarrow \Pi(p^2) - \Pi\left(\mu^2\right)$.  After 
regularization,
\beq
\Pi(p^2) = - \frac{1}{2\pi^{2}} \int_0^1 d\;x \; x\left(1 -x \right)
{\rm log} \left[ \frac{m_{UV}^2 + x\left(1 - x\right) \mu^2}
{m_{UV}^2 + x\left(1 - x\right) p^2} \right]~.
\eeq

Renormalization group flow is obtained by varying the scale $\mu$ 
above.  The physical amplitudes remain invariant if the coefficient of 
the brane localized kinetic term $1/g_{UV}^2$ is varied so as to keep 
the complete two point correlator invariant.  We obtain the following 
renormalization group equation for $1/g_{UV}^2$,
\begin{equation}
\frac{d \; g_{UV}}{d \; {\rm log} \mu} = \frac{g_{UV}^3}{2 \pi^2}
\int_0^1 dx \; \mu^2 \frac{x^2\left(1 - x\right)^2}{m_{UV}^2 + 
x\left(1 - x\right) \mu^2}
\end{equation}
This renormalization group equation is valid from the scale $M_5$ down to the
compactification scale.
For $\mu^2 \gg m_{UV}^2$ we see that the renormalization group 
evolution of the brane localized kinetic term is the familiar one for 
QED in four dimensions, while for $\mu^2 \ll m_{UV}^2$ the beta 
function tends to zero.  Thus in this mass dependent renormalization 
scheme the decoupling of heavy modes is occurring automatically.  This 
is the primary advantage of this scheme. 

We now turn to the corrections to the two-point correlator from the IR 
brane.  The fermionic loop on the IR brane leads to
\begin{equation}
G_{\mu \nu}\left(0,0;p\right) \rightarrow
G_{\mu \nu}\left(0, 0 ;p\right) +
G_{\mu \alpha}\left(0,\LL ;p\right)\left[p^2 \eta^{\alpha \beta} - 
p^{\alpha} p^{\beta}
\right] i \Pi(p^2) G_{\beta \nu}\left(\LL,0;p\right)~.
\end{equation}
Once again $\Pi(p^2)$ must be regulated.  As before we
regulate this by momentum subtraction at a scale $p^2 = \mu^2$.  
After regularization,
\begin{equation}
\Pi(p^2) = - \frac{1}{2\pi^{2}} \int_0^1 d\;x \; x\left(1 -x \right)
{\rm log} \left[ \frac{m_{IR}^2 \; e^{-2k \LL} + x\left(1 - x\right) \mu^2}
{m_{IR}^2 \; e^{-2k \LL} + x\left(1 - x\right) p^2} \right]~.
\end{equation}

The requirement that the two point function be independent of the 
arbitrary scale $\mu$ leads once again to a renormalization group 
equation for the coefficient of the brane localized kinetic term on 
the IR brane.
\begin{equation}
\frac{d \; g_{IR}}{d \; {\rm log} \mu} = \frac{g_{IR}^3}{2 \pi^2}
\int_0^1 dx \; \mu^2 \frac{x^2\left(1 - x\right)^2}{m_{IR}^2 \; e^{-2k \LL}
 + x\left(1 - x\right) \mu^2}
\end{equation}

This renormalization group equation is valid from the cutoff on the IR 
brane $M_5\,e^{-k\LL}$ down to the compactification scale.  We see 
that in this case the fermion decouples at a scale $m_{IR}\,e^{-k 
\LL}$.  Far above this scale the beta function is the one familiar 
from QED, while far below this scale it tends to zero.

The four dimensional effective theory for this model is obtained by 
performing a Kaluza Klein decomposition of the higher dimensional 
theory and integrating over the extra dimension.  Provided the 
renormalization scale $\mu$ at which the entire higher dimensional 
theory is defined is close to the compactification scale then tree 
level matching of the parameters will be accurate up to loop effects 
which are not logarithmically enhanced.

It is important to know whether the framework we have described above 
can be generalized to the cases when there are bulk fields 
transforming under the gauge group, and when the gauge group is 
non-Abelian.  Goldberger and Rothstein {\cite{Goldberger:2002cz}} have 
shown that the parameters of a U(1) theory with a bulk scalar field 
charged under the gauge group can be renormalization group evolved 
using correlators with external legs on the UV brane.  However 
non-Abelian gauge theories have yet to be examined using this 
framework.  Nevertheless in the sections that follow we will assume 
that it is possible to do this and proceed.

Now consider a `GUT' model consisting of a U(1)$_A \times $ U(1)$_B$ 
gauge theory.  Each U(1) has a Dirac fermion transforming under it on 
each of the two branes.  The GUT symmetry here is a $Z_2$ which 
interchanges the gauge fields and the fermions transforming under 
them.  What we are interested in understanding is the effect on the 
parameters on the IR brane if the GUT symmetry is broken on the UV 
brane.  This will be useful for our purposes later.  Suppose that the 
GUT symmetry is explicitly broken on the UV brane so that $m_{UV;A} > 
m_{UV;B}$.  However we see from the renormalization group equation for 
$g_{IR}$ that (at least to one loop order) $g_{IR;A} = g_{IR;B}$ even 
below the scale $m_{UV;A}$.  However below this scale $g_{UV;A} \neq 
g_{UV;B}$ so that the U(1) gauge coupling strengths in the four 
dimensional effective theory will be different.  Nevertheless we see 
from this that parameters on the IR brane are screened from effects of 
symmetry breaking on the UV brane above the compactification scale.

\subsection{Effect of Localized Gauge Symmetry Breaking}

Now we consider the effects of gauge symmetry breaking.  We will 
restrict to the case where the breaking is due to the VEV of a Higgs 
field localized on the UV brane.  Since we are interested in the case 
where the gauge symmetry is broken at a high scale $v \sim M_{GUT}\sim 
10^{16}~ \rm{GeV}$, the effects of $v$ must be included exactly in the 
propagator at tree-level.  We leave the relevant details to appendix 
A, where the exact tree-level gauge boson propagator in $A_5 = 0$ 
gauge and four-dimensional unitary gauge is calculated.  There, we 
also summarize the asymptotic forms of the resulting propagator in the 
various energy regimes.  We base the present discussion on the 
asymptotic behavior of the propagator, which is physically more 
transparent.  Referring to the propagator $G_p(y,y)$ defined in 
Eq.~(\ref{Gmunu}) of appendix A, we find that for the broken gauge 
bosons, and provided $g_5^2 v^2 \lsim k$, its high energy behavior, $p 
\gg k\,e^{-k y}$, is unchanged compared to the unbroken case up to very 
small corrections (see Eqs.~(\ref{veryhighp}) and (\ref{highp}) of 
appendix A).  Thus, if the VEV $v$ is responsible for the breaking of 
the gauge group G down to a subgroup H, observers at $y$ will see (at 
least at tree-level) the full symmetry G when probing energies above 
$k\,e^{-k y}$, a scale that is in general much lower than the naive 
scale of symmetry breaking of order $v$.  Only for momenta $p \ll 
k\,e^{-k y}$ do observers at $y$ have a chance to see that the gauge 
symmetry is actually broken.  In that energy regime and without making 
any assumptions about the size of $v$, we find (see Eqs.~(\ref{lowp}) 
and (\ref{verylowp}) of appendix A)
\beq
\label{lowenergypropagator}
G_p(y,y) \sim - \frac{g_5^2}{2} \;
\frac{2 k+(e^{2 k y}-1) \, g_5^2 v^2}{p^2 \ln(2k/p) + k \, g_5^2 v^2}
\hspace{1cm} k\,e^{-k\LL} \ll p \ll k\,e^{-ky}~,
\eeq
while
\beq
\label{verylowenergypropagator}
G_p(y,y) \sim - \frac{g_5^2}{2 \, k} \;
\frac{2 k+(e^{2 k y}-1) \, g_5^2 v^2}{ \LL p^2 + \, g_5^2 v^2}
\hspace{1cm} p \ll k\,e^{-k\LL}~.
\eeq
We see again that Eq.~(\ref{verylowenergypropagator}) can be obtained from
Eq.~(\ref{lowenergypropagator}) by making the replacement $\ln(2k/p)
\rightarrow k \LL$.

Specializing to the case of interest here, where $v \sim M_{\rm{GUT}}
\sim k$, we note from Eq.~(\ref{verylowenergypropagator}) that at very
low energies, $p \ll k\,e^{-k \LL}$, the UV propagator is
\beq
G_p(0,0) \sim - \frac{1}{v^{2}}~.
\eeq
Therefore, when matter is localized on the UV brane, processes 
mediated by the broken gauge bosons, such as proton decay, are exactly 
as suppressed as in conventional four-dimensional theories.  For IR 
localized observers, one finds instead
\beq
G_p(L,L) \sim - \frac{g_5^2}{2k} \, e^{2 k \LL}
\sim - \frac{g_4^2}{2k/\LL} \, e^{2 k \LL}~,
\eeq
where $g_{4}^{2} = g_{5}^{2}/L$ is the four-dimensional gauge coupling 
of the unbroken gauge bosons.  Thus, the broken gauge bosons appear to 
have a mass of order $(k/\LL)^{1/2}\,e^{-k \LL}$ from the point of 
view of IR observers.  This is consistent with our previous 
observation that the broken gauge bosons behave like massless fields 
above the compactification scale $k\,e^{-k \LL}$.

The picture suggested by the above discussion is that all IR 
correlators will exhibit relations appropriate to the full symmetry 
group of the theory at energies somewhat larger than the 
compactification scale $k\,e^{-k \LL}$.  The same would hold for the 
coefficients of operators localized on the IR brane.  In order to make 
the above statement precise one has to go beyond the previous 
tree-level analysis and argue that loop effects will not modify this 
picture.  We will consider this in a subsequent section.  Nevertheless 
this is completely consistent with our results for the case of the 
discrete $Z_2$ symmetry in the previous subsection.

\section{Localized Supersymmetry Breaking}

We are considering a scenario where supersymmetry is broken on the 
infrared brane and supersymmetry breaking effects are transmitted to 
fields localized on the ultraviolet brane by the gauge multiplet in 
the bulk.  However, the GUT symmetry is assumed to be broken on the 
ultraviolet brane, at a scale $M_{GUT}$ much higher than the 
compactification scale, which we denote by $M_c \equiv k\,e^{-k\LL}$.  
We must therefore be careful to take GUT symmetry breaking effects 
into account when computing the sparticle masses.  The hierarchy of 
scales we are considering is $M_5 \gsim k \gsim M_{GUT} \gg M_c, 
\sqrt{F}$.  At energy scales well above the GUT scale the Planck brane 
correlators of the theory will respect the GUT symmetry.  Hence the 
parameters of the bulk Lagrangian and the parameters of the Lagrangian 
on the UV brane defined at a 4-d momentum scale $\mu > M_{GUT}$ will 
respect the GUT symmetry, and any divergences that arise at loop level 
can be removed by adding counterterms that respect this symmetry.  
However the parameters on the IR brane cannot be defined at a scale so 
much higher than the cutoff on the IR brane.  Nevertheless the results 
of the previous section suggest that they can consistently be defined 
as GUT symmetric at a scale close to the cutoff on the IR brane 
$M_5\,e^{-k \LL}$.  This is because physics on the IR brane is 
screened from the effects of GUT symmetry breaking on the UV brane 
above the compactification scale.

At 4-d momentum scales lower than the GUT scale the Planck brane 
correlators will in general no longer respect the GUT symmetry, 
corresponding to the fact that the parameters of the Lagrangian on 
the UV brane are no longer GUT symmetric.  Our approach will be to 
evolve the higher dimensional Lagrangian from the GUT scale to a scale 
$\mu \approx M_c$ and then match on to the four dimensional
effective theory at tree level.  Any matching corrections to our 
expressions will then be loop suppressed and in particular, they will 
not be enhanced by the logarithm of $M_{GUT}/M_c$.

We begin by investigating the effects of GUT symmetry breaking on the
supersymmetric terms in the Lagrangian.  Consider the form of the 
Lagrangian for the gauginos.  At 4-d scales above the GUT scale the 
supersymmetric part of the action has a bulk contribution
\beq
\label{eq:bulkgauge}
-\frac{1}{g_5^2} \int_0^\LL d^4\!x\,dy \sqrt{-G} \;
\sum_{A \in SU(5)} \left[
\frac{i}{2} \bar{\lambda}^{iA} \Gamma^M D_M \lambda^{iA} +
\frac{i}{4} \sigma' \bar{\lambda}^{iA} \left( \sigma_3 \right)^{ij} \lambda^{jA}
\right]~,
\eeq
and a brane localized contribution
\beq
\label{eq:branegauge1}
-\frac{1}{g_{UV}^2} \int_0^\LL d^4\!x\,dy \sqrt{-G_{\rm{ind}}}\;
2\delta\left(y \right)
\sum_{A \in SU(5)}
\left[ 
i \bar{\lambda}^{1A} \Gamma^{\mu} D_{\mu} P_L \lambda^{1A}
\right]
\eeq
on the UV brane, where we inserted the chirality projector $P_L$ so
that only the left-handed components of the bulk gaugino appear in
the brane term.

At energy scales below the GUT scale the form of the action changes.
While the forms of the bulk action Eq.~(\ref{eq:bulkgauge}) and the
IR brane action Eq.~(\ref{eq:branegauge2}) remain the same, the
contribution to the action from the UV brane where the GUT symmetry is
broken now takes the form
\beq
\label{eq:branegauge3}
-\int_0^\LL d^4\!x\,dy \sqrt{-G_{\rm{ind}}} \;
2\delta\left(y - \LL \right) \sum_{A \in SU(5)} \left[
\frac{1}{g_{UV, A}^2}\,i\,
 \bar{\lambda}^{1A} \Gamma^{\mu} D_{\mu}
P_L \lambda^{1A} \right]~,
\eeq
where because of GUT symmetry breaking $g_{UV,A}^{2}$ is in general not 
the same for SU(3), SU(2), U(1) and SU(5)/[SU$(3) \times$
SU(2)$\times$U(1)].

The kinetic terms for the gaugino on the IR brane are defined at the scale 
 $M_5\,e^{-k\LL}$ and take the SU(5) symmetric form
\beq \label{eq:branegauge2} 
-\frac{1}{g_{IR}^2} \int_0^\LL d^4\!x\,dy \sqrt{-G_{\rm{ind}}}\;
2\delta\left(y-\LL\right) 
\sum_{A \in SU(5)}\,i\,\bar{\lambda}^{1A} \Gamma^{\mu}
D_{\mu} P_L \lambda^{1A}
\eeq
Renormalization group evolving this down to $M_c$ does not alter the 
SU(5) symmetric form.

In order to obtain the four dimensional effective action we determine 
the form of the complete higher dimensional action at a scale close to 
$M_c$ and after performing a Kaluza- Klein decomposition we then 
integrate over the extra dimension.  Since the SU(5)/[SU$(3) 
\times$SU(2)$\times$U(1)] gauginos all have masses of order the 
compactification scale or higher the only surviving degrees of freedom 
are the zero modes of the SU(3), SU(2) and U(1) gauginos.  The four 
dimensional effective theory for the zero mode gauginos is now the one 
familiar from four dimensions:
\beq
\label{4Dgauginoaction}
-\int d^4 x \sqrt{-g}
\sum_{A \in SU(3),SU(2),U(1)}\frac{1}{g_{A}^2}\,\frac{i}{2}\, 
\bar{\lambda}_{L}^{A} {\Gamma}^{\mu}{D}_{\mu} \lambda_{L}^{A}~,
\eeq
where now $\lambda_{L}^{A} = (\lambda^{1}_{\alpha}, 
\bar{\lambda}^{1\dot{\alpha}})$ is a four component Majorana spinor 
built from the even components of the original symplectic Majorana 
pair (all other components get masses of the order of the 
compactification scale). Also, the relation between the 4-dimensional
gauge coupling $g_A^2$ and the fundamental parameters in the theory
takes the form
\beq
\frac{1}{g_A^2} = \frac{\LL}{g_5^2} + \frac{1}{g_{IR}^2} +
\frac{1}{g_{UV,A}^2}~,
\eeq
where the first two terms are essentially GUT symmetric, but the last
one contains nonuniversal logarithms from loop effects.

We now turn our attention to supersymmetry breaking terms.  Consider
the effect of the following supersymmetry breaking term localized on
the infrared brane.
\beq
\int_0^\LL d^4\!x\,dy \sqrt{-G_{\rm{ind}}} \; 
2\delta \left( y - \LL \right)  
\sum_{A \in SU(5)}
\left\{\frac{1}{8}\int d^4 \theta \frac{X^{\dagger} X}{M_5^4}
W^{\alpha A} D^2 W_{\alpha}^{A} + H.c. \right\}~, 
\eeq 
where $X$ is the hidden sector superfield that breaks supersymmetry
($F_X \neq 0$) and $W_{\alpha}^{A} = - i \lambda_{\alpha}^{1A} + 
\cdots$ is the field strength superfield constructed from the
left-handed component of $\lambda^{1A}$.  We follow the conventions of 
\cite{WessBagger}.  The effect of such a term is to alter the relative 
strengths of the couplings of the gauge boson and gaugino to 
matter{\footnote{The implications of such terms for extra 
dimensional supersymmetry
breaking were first pointed out in 
{\cite{Kaplan:2000jz}}.}}. This contributes to the zero mode gaugino 
Lagrangian a term
\beq
\label{gauginoassisted} 
-\int_0^\LL d^4\!x\,dy \sqrt{-G_{\rm{ind}}} \; 2\delta \left( y - \LL
\right) \sum_{A \in SU(5)} \frac{|F|^{2}}{M_{5}^{4}}\,\,\frac{i}{2}\,
\bar{\lambda}_{L}^{A} \Gamma^\mu {D}_\mu \lambda_{L}^{A}~, 
\eeq 
where the four-component spinors $\lambda_{L}^{A}$ are defined as in 
Eq.~(\ref{4Dgauginoaction}), but keeping the full $y$-dependence.  The 
arguments of the previous section suggest that this term is SU(5) 
symmetric at compactification scale $k\,e^{-k\LL}$.  We now show that 
if the GUT symmetry is only broken on the UV brane loop corrections to 
this term from above the compactification scale do indeed respect the 
SU(5) symmetric form.

To determine the form of this term at the matching scale $M_c$ we 
consider the two point function for the gaugino in mixed position 
momentum space with four momentum of order the IR scale and the end 
points on the UV brane.  We base our discussion in terms of the 
properties of the tree-level gaugino propagator.  For example, we find 
that the propagator for the even spinor components, $G_{LL}^{A} \equiv 
P_L G^{A} P_R = \langle \lambda_L^{A} \overline{\lambda_L^{A}} 
\rangle$, is
\beq
\label{GLL}
G_{LL}^{A}(y,y';p) = - i P_L \Dslash{p}\, e^{\frac{3}{2}k(y+y')} 
G_p^{A}(y,y')\hspace{1cm}A \in SU(5)~,
\eeq
where $G_p^{A}(y,y')$ is precisely the scalar part of the gauge boson 
propagator (see Eq~(\ref{GaugeBosonPropagator}) in appendix A), whose 
properties where discussed in Section 3.  Equation~(\ref{GLL}) holds 
for the unbroken as well as for the broken fields.  We provide the 
details of the derivation of the gaugino propagator in the presence of 
UV localized gauge symmetry breaking in appendix B, where we also give 
the exact expressions for $G_{RR}$, $G_{RL}$ and $G_{LR}$ (see 
Eqs.~(\ref{oddpropagator}), (\ref{GLR}) and (\ref{GRL})).

If we treat the operator Eq.~(\ref{gauginoassisted}) perturbatively,
the UV brane two-point gaugino correlator receives an SU(5)
symmetric tree-level contribution proportional to
\beq
G^{A}_{LL}\left( 0,\LL;p\right) \left(\Gamma^{\alpha} p_{\alpha}\,
\frac{|F|^{2}}{M_{5}^{4}}\,e^{-3 k \LL}\right)
G^{A}_{LL}\left(\LL,0;p\right)~,
\eeq
where the factor $e^{-3 k \LL}$ comes from the determinant of the 
induced metric and the vielbein implicit in the contraction 
$\Gamma^{\mu}D_{\mu}$ (see Fig.~1~a).  Now consider loop corrections 
with one insertion of Eq.~(\ref{gauginoassisted}).  One of the 
contributions to the leading one-loop amplitude is shown in Fig.~1~b 
and has the form
\beqa
\sum_{B,C \in SU(5)} \left(f^{ABC}\right)^2  
\int \frac{d^4 q}{(2\pi)^{4}} \int_{0}^{L} dy' \int _{0}^{L} dy'' 
G^{A}_{LL}\left( 0,y';p\right) \Gamma^{\mu} 
G_{LL}^{C}\left( y',\LL;p+q\right)
\\ \nonumber
\mbox{} \times \left(\Gamma^{\alpha} p_{\alpha}\, 
\frac{|F|^{2}}{M_{5}^{4}}\,e^{-3 k \LL} \right)
G_{LL}^{C}\left( \LL,y'';p+q\right)
G^{B}_{\mu \nu}\left( y', y'';q\right)
\Gamma^{\nu} G^{A}_{LL}\left( y'',0;p\right)~,
\eeqa
where $f^{ABC}$ are the structure constants of $SU(5)$.

\begin{figure}[t]
\vspace*{-1.cm}
\centerline{ 
   \resizebox{6cm}{!}{\includegraphics{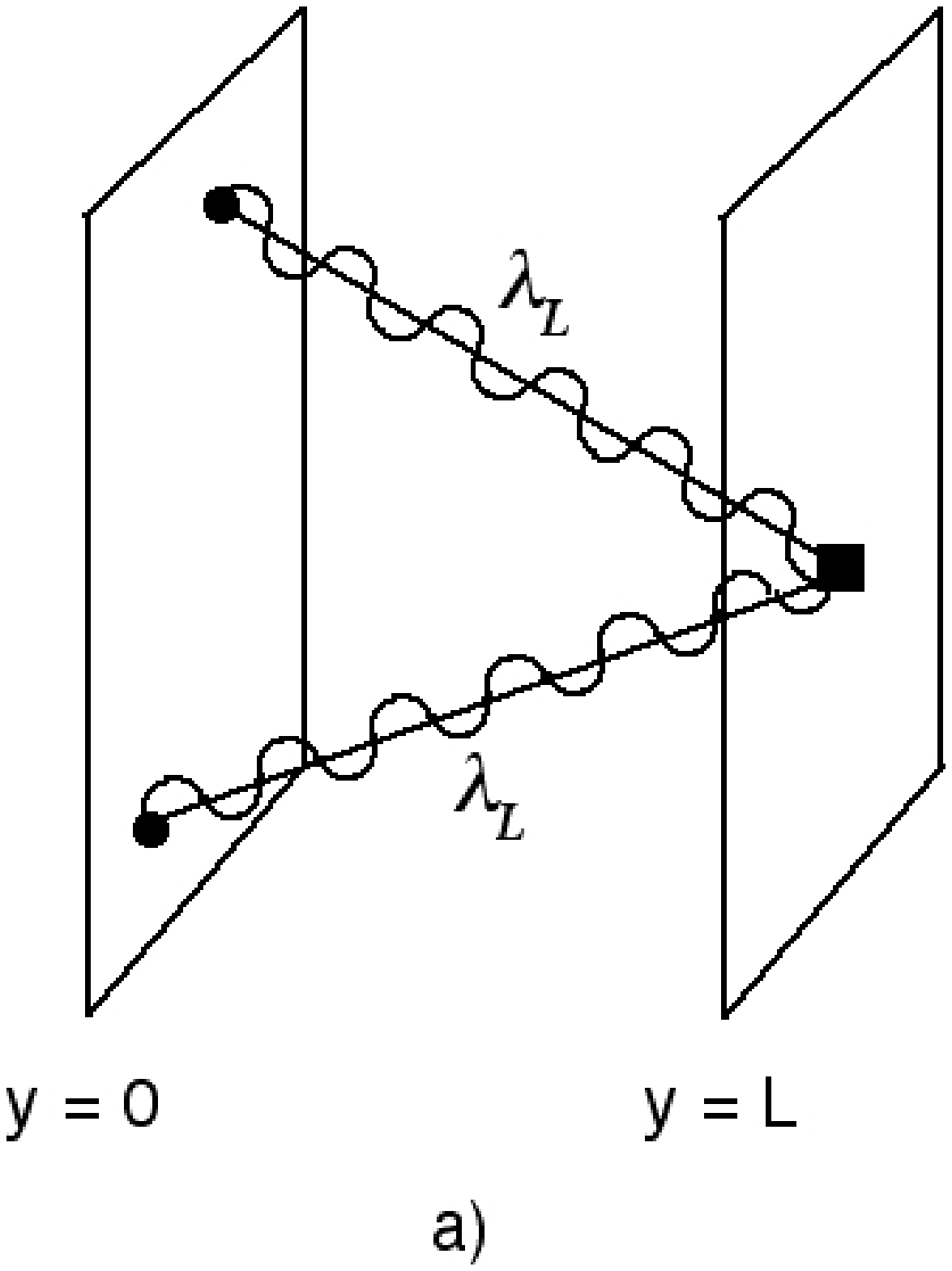}} \hspace*{2cm}
   \resizebox{6cm}{!}{\includegraphics{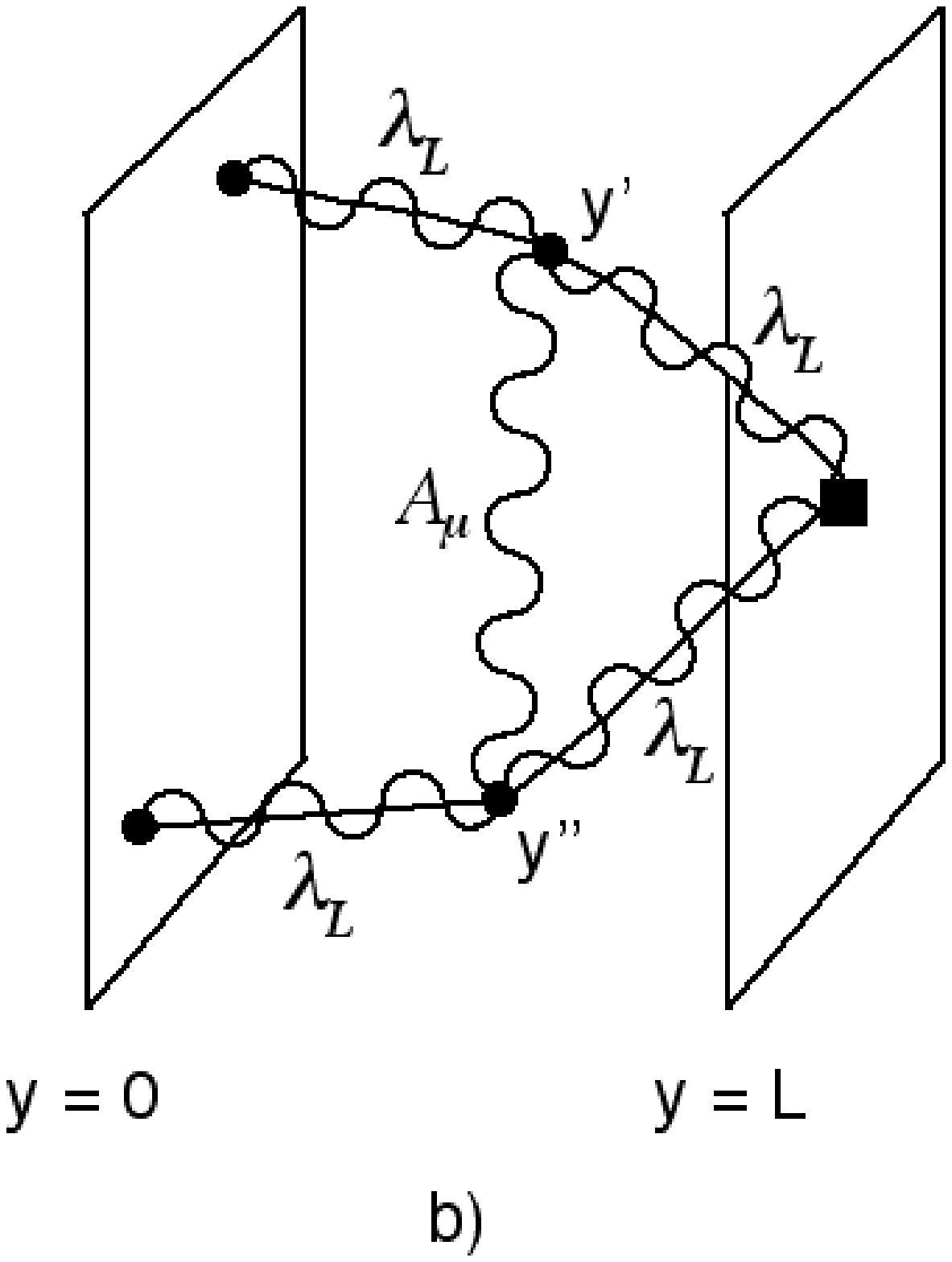}}
}
\caption{Contributions to the two-point gaugino correlator with end
points on the UV brane from the operator Eq.~(\ref{gauginoassisted}),
represented by a square, in the one insertion approximation. a)
Tree-level and b) one of the 1-loop contributions.}
\label{fig:twopoint}
\end{figure}

This diagram is divergent for large loop momenta $q$ when the vertices 
$y'$ and $y''$ become close to $\LL$.  However the form of this 
divergence is SU(5) symmetric, since the forms of the relevant
propagators for both the broken and unbroken generators are the same 
up to exponentially small differences when $q \gg k\,e^{-k \LL}$.  
From the forms of the propagators it can be further inferred that 
SU(5) violating corrections to the supersymmetry breaking term
Eq.~(\ref{gauginoassisted}), if any, from loop momenta much higher 
than the IR scale are exponentially suppressed and therefore small.  
From this we conclude that the form of this parameter in the higher 
dimensional theory at a scale close to the matching scale is SU(5)
symmetric and any corrections to this are loop suppressed and without 
any logarithmic enhancement.  Based on this we argue that any 
supersymmetry breaking term localized on the IR brane will have an 
SU(5) symmetric form up to loop suppressed corrections at four 
dimensional momenta of order the compactification scale.

We now obtain an expression for the gaugino soft masses in the limit
where the SUSY breaking scale is well below the compactification
scale $M_c$. These arise from operators of the form
\beq
\label{gauginooperator}
\int_0^\LL d^4\!x\,dy \sqrt{-G_{\rm{ind}}} \;
2\delta \left( y - \LL \right) \sum_{A \in SU(5)} 
\frac{1}{2} \int d^2 \theta \frac{\lambda X}{M_5} 
W^{\alpha A} W_{\alpha}^{A}~,
\eeq
localized on the hidden brane where supersymmetry is broken.  Here $X$ 
is the hidden sector field which breaks supersymmetry, $\lambda$ is a 
coupling constant, and we are assuming that $F_X \neq 0$.  From our 
previous discussion, this term is SU(5) symmetric at a scale $\mu$ of 
order the compactification scale, $M_c$.  We now do a Kaluza Klein 
decomposition and integrate out the massive KK modes at tree-level.  
We are assuming that the effective SUSY breaking VEV, $\sqrt{F} e^{-k 
\LL}$ is small compared to the compactification scale $M_c$, so 
that it can be treated perturbatively.  We find then from the 
resulting action for the zero mode that the physical gaugino mass, 
which arises from Eq.~(\ref{gauginooperator}) after canonically 
normalizing the gaugino kinetic term, is (see appendix C)
\beq
\label{gauginomass}
m_{\lambda_{i}}(\mu = M_{c}) = g_i^2(\mu = M_c)
\left[\frac{\lambda F}{M_5}\, e^{-k \LL}\right]
\hspace{1cm} i = 1,2,3~.
\eeq
The single power of the warp factor $e^{-k\LL}$ arises from the
determinant of the induced metric and the zero-mode wavefunction given 
in Eq.~(\ref{gauginozeromode}).  We see that the gaugino mass is 
proportional to the square of the low energy gauge coupling just as in 
conventional gaugino mediation, but is scaled down by the warp factor 
reflecting the fact that it arises from a contact interaction on the
IR brane.

We now turn to the scalar masses.  Due to the sequestering, direct 
contact terms between the supersymmetry breaking sector, localized on 
the IR brane, and the observable sector, localized on the UV brane,
are exponentially suppressed.  A larger source of scalar masses arises 
from loop diagrams involving the soft supersymmetry breaking gaugino 
masses, just as in conventional gaugino mediation.  Contributions to 
the soft scalar masses from 4-d loop momenta larger than the 
compactification scale are exponentially suppressed because of the 
form of the gaugino propagator $G_{LL}(0,\LL;p)$, which in this energy 
regime behaves like $e^{-p e^{k \LL}/k}$ (see Eqs.~(\ref{GLL}), 
(\ref{veryhighp}), (\ref{highp}) and (\ref{nonlocal})).  Hence if the 
compactification scale is much higher than the weak scale the dominant 
contribution to the scalar masses arises from renormalization group 
evolving the scalar masses from the compactification scale down to the 
weak scale.  The scalar masses are loop suppressed but enhanced by the 
logarithm of $M_c/M_{\rm{WEAK}}$ with respect to the gaugino masses.  
However if the compactification scale is 
not hierarchically larger than the weak scale then 
the matching corrections from loop momenta close to the 
compactification scale must be taken into account since the logarithm 
of $M_c/M_{\rm{WEAK}}$ is not large enough to justify neglecting them.

In the limit where the compactication scale is close to the weak scale 
the spectrum can differ significantly from that of the theory with 
high compactification scale.  The effects of the repulsion of the 
gaugino wave function from the IR brane \cite{Chacko:1999hg, 
Arkani-Hamed:2001mi} can now no longer be neglected and must be taken 
into account.  While the spectrum of gaugino masses becomes more 
complicated, it remains hierarchical and in the limit of strong 
supersymmetry breaking $g_{5}^{2} \lambda F/M_{5} \gg \sqrt{k\LL}$ the 
lightest gaugino is a pseudo-Dirac state which has a mass given by 
(see Eq.~(\ref{largeSUSY}))
\beq
\label{strongSUSY}
m_i \approx \sqrt{2 k} \, \frac{g_i(\mu)}{g_5}\,e^{-k\LL}~,
\eeq
where $\mu \sim M_c$, and 
\beq
\label{gauginogaugecoupling}
\frac{1}{g_i^2(\mu)} = \frac{\LL}{g_5^2} + \frac{1}{g_{UV,i}^2(\mu)}~.
\eeq
This result is derived in appendix C. When the IR localized kinetic 
terms can be neglected, $g_{i}$ coincides with the low-energy gauge 
coupling and therefore, for the SM, the two terms on the rhs in 
Eq.~(\ref{gauginogaugecoupling}) are of the same order.  We note that 
in this limit the gaugino masses are independent of the SUSY breaking 
parameter $F$, and are proportional to $g_i$, instead of $g_i^2$ as is 
the case when the SUSY scale is much smaller than the compactification 
scale.  It is also important that the gauginos corresponding to the 
different gauge groups are {\textit{not}} degenerate.

Equation (\ref{strongSUSY}) differs substantially from the standard 
gaugino mediation formula and shows that, in the case where the 
compactification scale is close to the weak scale, the gaugino masses 
are approximately in the ratio $M_3$:$M_2$:$M_1$ : 2.3:1.4:1.  In this 
scenario the gauginos will be in the few TeV range.  

The scalar masses arise from finite 1-loop diagrams.  However, in the 
limit that the compactification scale is close to the weak scale, they 
depend on the coefficients of operators like 
Eq.~(\ref{gauginoassisted}) and thus depend on more parameters.  As 
mentioned before, operators of this form have the effect of altering 
the relative strengths of the couplings of gauge bosons and gauginos 
to matter and therefore constitute a hard breaking of supersymmetry.  
They cannot be forbidden by any symmetry.  The scalar masses squared 
receive contributions from loop diagrams involving gauginos with 
insertions of this operator.  These diagrams are cutoff at the 
compactification scale because of the exponential fall-off of the 
gaugino propagator for 4-d momenta larger than the compactification 
scale.  Simple scaling suggests that the ratio of the gaugino mediated 
contribution to the contribution from hard breaking operators in the 
limit that $M_c \approx M_{\rm{WEAK}}$ is of order 
$\left(M_5/k\right)^2$.  This shows that the effects of the hard 
supersymmetry breaking operators on the scalar masses can be safely 
neglected only if the logarithm of $M_c/M_{\rm{WEAK}}$ is much larger 
than one or if $M_5$ is significantly larger than $k$.

\section{Phenomenology}

We now turn to the phenomenology of the present class of models.  We 
concentrate on the case where $M_c \gsim 10^{6}~\rm{GeV}$, so that the 
contributions from operators like Eq.~(\ref{gauginoassisted}) can be 
neglected.  We have established that, under the assumption of grand 
unification, and provided supersymmetry is broken at a scale much 
lower than the compactification scale $k\, e^{-k L}$, the low energy 
gaugino masses are given by \beq m_i(\mu) = g_i^2(\mu) M_{SUSY}~, \eeq 
where $M_{SUSY} = (\lambda F/M_{5}) e^{-k \LL}$ sets the scale for 
supersymmetry breaking in the observable sector and is common to the 
$SU(3)$, $SU(2)_{L}$ and $U(1)_{Y}$ gauginos.  This formula is very 
similar to the one obtained in the case where the extra dimension is 
flat and sufficiently small that unification takes place within an 
effective four-dimensional theory.  Scalar masses are also induced at 
the compactification scale $M_c = k\, e^{-k \LL}$, but are one-loop 
suppressed compared to the gaugino masses.

In order to relate these mass parameters to the physical superpartner 
masses, it is necessary to evolve them via the renormalization group 
equations from the compactification scale $M_c=k\, e^{-k \LL}$ to the 
weak scale, which we take as $1~\rm{TeV}$.  Since $M_c$ can easily be 
much smaller than the GUT scale, the resulting superpartner spectrum 
will be, in general, very different from the one obtained when the 
extra dimension is flat.  When $M_{c}$ is not close to the weak scale, 
we can neglect the matching contributions to the scalar masses at the 
compactification scale, compared to the logarithmically enhanced 
contributions from the running below $M_c$.  Thus, we may take as the 
initial conditions in the solution to the RG equations 
$m_{s}^{2}(\mu=M_c) \simeq 0$, for all scalars.  It is well-known that 
the scalar squared masses generated through the RG evolution are 
positive.  The only exception is the up-type Higgs mass$^{2}$, which 
is driven negative by the large top Yukawa coupling, thus triggering 
electroweak symmetry breaking.

The minimization of the Higgs potential requires the specification of 
the $\mu$ and $B_{\mu}$ parameters.  We will take $\tan\beta$ as a 
free parameter and fix $\mu$ by requiring that the observed $M_{Z}$ is 
reproduced.  It is also important to include the radiative corrections 
to the Higgs potential.  The most important correction comes from 
top-stop loops, and can be taken into account by including the term
\beq
\Delta V_{H} = \frac{3 y_{t}^{4}}{16 \pi^{2}}
\ln\left[\frac{m_{\tilde{t}_{1}} m_{\tilde{t}_{2}}}{m_{t}^{2}}\right]
(H_{u}^{\dagger} H_{u})^{2}~,
\eeq
where $y_{t}$ is the top Yukawa coupling, $m_{\tilde{t}_{i}}$ are the 
physical masses of the stops, and $m_{t} \simeq 165~\rm{GeV}$ is the 
top mass (in the absence of QCD corrections).

\begin{table}
\centering
\begin{tabular}{|c|c|c||c|c|}
\hline
&  & Point 1 & Point 2 & Point 3\\
\hline
inputs: & $M_c$ & $2 \times 10^{16}$ & $10^{11}$ & $10^{6}$ \\
& $M_{SUSY}$ & 344 & 318 & 473 \\
& $\tan\beta$ & 10 & 10 & 10 \\
\hline
neutralinos:
& $m_{\chi^0_1}$ & 148 & 208 & 406 \\
& $m_{\chi^0_2}$ & 272 & 381 & 556 \\
& $m_{\chi^0_3}$ & 468 & 535 & 566 \\
& $m_{\chi^0_4}$ & 487 & 560 & 808 \\
\hline
charginos: 
& $m_{\chi^\pm_1}$ & 272 & 380 & 550 \\
& $m_{\chi^\pm_2}$ & 487 & 559 & 808 \\
\hline
Higgs: 
& $m_{h^0}$ & 114 & 119 & 126 \\
& $m_{H^0}$ & 498 & 560 & 605 \\
& $m_A$ & 498 & 560 & 604 \\
& $m_{H^\pm}$ & 504 & 566 & 610 \\
\hline
sleptons: 
& $m_{\tilde{\nu}_L}$ & 231 & 232 & 267 \\
& $m_{\tilde{e}_L}$ & 245 & 246 & 278 \\
& $m_{\tilde{e}_R}$ & 139 & 111 & 108 \\
& $m_{\tilde{\tau}_2}$ & 247 & 250 & 281 \\
& $m_{\tilde{\tau}_1}$ & 130 & 100 & 100 \\
\hline
squarks: 
& $m_{\tilde{u}_L}$ & 745 & 945 & 1420 \\
& $m_{\tilde{d}_L}$ & 750 & 950 & 1420 \\
& $m_{\tilde{u}_R}$ & 715 & 920 & 1400 \\
& $m_{\tilde{d}_R}$ & 715 & 920 & 1400 \\
& $m_{\tilde{t}_2}$ & 735 & 935 & 1410 \\
& $m_{\tilde{t}_1}$ & 545 & 760 & 1280 \\
\hline
gluino: 
& $M_3$ & 830 & 1060 & 2280 \\
\hline
\end{tabular}
\capt{Sample points in parameter space.  All masses are in GeV. The 
first point corresponds to gaugino mediation in flat space.  The last 
two points reflect the effect of warping through a lower effective 
compactification scale $M_c=k\,e^{-k \LL}$. In these cases the LSP is 
the gravitino.}
\end{table}

The superpartner spectrum also depends on the scale of SUSY breaking, 
$M_{SUSY}$, which simply sets the overall scale for all the soft 
terms.  Therefore, the model has three independent parameters: the 
compactification scale $M_c = k \, e^{-k \LL}$, a common gaugino mass 
parameter $M_{SUSY}$ and $\tan\beta$.  We obtain all other 
superparticle masses by integrating numerically the one-loop RG 
equations.  In Table 1, we give some sample points for $\tan\beta = 
10$, and different values of the compactification scale: $M_c = 
M_{GUT}$ (``standard'' gaugino mediation), an intermediate scale $M_c 
= 10^{11}~\rm{GeV}$ and a low scale $M_c = 10^{6}~\rm{GeV}$.  
$M_{SUSY}$ is chosen so as to satisfy the experimental bounds.  For 
the first point in the table, the strongest constraint comes from 
requiring that the lightest Higgs mass satisfies $m_{h} > 
114~\rm{GeV}$.  For the last two points, the strongest constraint 
comes from requiring that the right-handed sleptons be above 
$100~\rm{GeV}$.  In these cases, the lightest Higgs mass is found to 
be $119~\rm{GeV}$ and $126~\rm{GeV}$, respectively.

A distinct feature of the present class of models is that the 
neutralinos are rather heavy.  This is due to the fact that the 
right-handed sleptons get their masses at loop-level only from the 
$U(1)_{Y}$ interactions.  The logarithmic enhancement from the 
renormalization group running is, in general, not enough to make them 
heavier than the bino, and it is necessary to take the overall scale 
of supersymmetry breaking, $M_{SUSY}$, to be in the few hundred GeV 
region.  Clearly, the hierarchy between the scalar and gaugino masses 
increases as the compactification scale is reduced, as illustrated in 
Table 1.  We should remark that the right-handed sleptons could get 
extra positive contributions if, for example, the Higgses propagate in 
the bulk of the extra dimension and pick a tree-level soft mass from 
direct interactions with the SUSY breaking sector.  In that case, the 
$\beta$-functions for the scalar masses receive a contribution 
proportional to
\beq
\xi = \sum_{i} Y_{i} m_{i}^{2}~,
\eeq
where the sum runs over all fields that have $U(1)_{Y}$ interactions 
and $Y_{i}$ is the hypercharge of the i-th field.  If $m_{H_{u}}^{2} > 
m_{H_{d}}^{2}$, this term gives a positive contribution to the 
right-handed sleptons.  However, for a low supersymmetry breaking 
scale, we do not expect this extra contribution to be enough to make 
the sleptons heavier than the lightest neutralino. 

Due to the Yukawa interactions, the three generations of sleptons are 
not degenerate.  The most important effect comes from the mixing term 
$\mu\,m_{l}\tan\beta$ between the right- and left-handed sleptons, 
where $m_{l}$ is the mass of the associated lepton (A-terms are very 
small in these models and there is no chance of a cancellation).  The 
result is to make the lightest superpartner mass lighter.  Although 
the effect in the smuon and selectron systems is small and can usually 
be neglected, in the stau system it can be quite important.  Thus, the 
next-to lightest-supersymmetric particle (NLSP) is the (mostly 
right-handed) stau, $\tilde{\tau}_{1}$.  (The LSP is alway the 
gravitino; see Eq.~(\ref{gravitinomass}) below.)  The collider 
phenomenology is similar to that of models with a low scale of SUSY 
breaking and a charged NLSP \cite{Giudice:1998bp}.  In particular, the 
termination of superpartner decay chains strongly depends on 
$\tan\beta$.  When $\tan\beta$ is not too large, the decay channels 
$\tilde{\mu}_{R} \rightarrow \mu \tau \tilde{\tau}_{1}$ and 
$\tilde{e}_{R} \rightarrow e \tau \tilde{\tau}_{1}$ are not 
kinematically open.  Then $\tilde{\mu}_{R}$ and $\tilde{e}_{R}$ decay 
predominantly into a gravitino and the corresponding lepton, giving 
rise to a slepton co-NLSP's scenario.\footnote{Competing three-body 
decays $\tilde{l} \rightarrow \nu_{l} \bar{\nu}_{\tau} 
\tilde{\tau}_{1}$, where $l=\mu$, $e$, through off-shell charginos are 
very suppressed because of phase space and because the couplings of 
the lightest sleptons to charginos is very small 
\cite{Ambrosanio:1997bq}.  We also assume that there is no R-parity 
violation.} For larger $\tan\beta$, the decays $\tilde{\mu}_{R} 
\rightarrow \mu \tau \tilde{\tau}_{1}$ and $\tilde{e}_{R} \rightarrow 
e \tau \tilde{\tau}_{1}$ are allowed and proceed through an off-shell 
neutralino, which in our case is always mostly bino.  Here it is 
important to distinguish among the $\tau^{+} \tilde{\tau}^{-}$ and 
$\tau^{-} \tilde{\tau}^{+}$ final states.  In fact, since the 
neutralinos are heavy, the equal-slepton-charge channel is suppressed 
compared to the opposite-slepton-charge one, as discussed in 
\cite{Ambrosanio:1997bq}.  The corresponding decay lengths increase 
with decreasing $\tan\beta$ and increasing neutralino mass.  For the 
last point in the table with $\tan\beta = 10$ and $m_{\tilde{N}_{1}} 
\simeq 400~GeV$, we estimate the decay length to be about $1~\rm{\mu 
m}$.

The subsequent decay of the $\tilde{\tau}_{1}$ into a gravitino is 
governed by the fundamental scale of SUSY breaking, $F$, as is the 
gravitino mass.  This scale is in turn related to $M_{SUSY}$ through a 
coupling $\lambda$ as in Eq.~(\ref{gauginooperator}).  We can estimate 
the size of this coupling by assuming that the theory becomes strongly 
coupled at a scale $\Lambda$, in which case an NDA estimate gives 
$M_{SUSY} = \tilde{F}/(l_{4}^{1/2} \tilde{\Lambda})$ 
\cite{Chacko:1999hg}.  Here a tilde is used to denote the 
corresponding parameters appropriately redshifted by the warp factor, 
and $l_{4}=16 \pi^{2}$.  Further using the NDA estimate $M_{5}^{3} = 
\Lambda^{3}/l_{5}$, with $l_{5}=24 \pi^{3}$, we can write
\beq 
\label{susyF}
\frac{F}{M_{P}^{2}} = l_{4}^{1/2} l_{5}^{1/3}
\left(\frac{k}{M_{5}}\right)^{2} \left(\frac{M_{SUSY}}{M_c}\right)~, 
\eeq 
where $M_{P} = 2 \times 10^{18}~\rm{GeV}$ is the reduced 
four-dimensional Planck scale as given in Eq.~(\ref{4DPlanckscale}) 
and $k/M_{5} \lsim 1$.  Taking $k/M_{5} = 1/10$ we find that for the 
sample points given in Table 1., the $\tilde{\tau}_{1}$ decays 
outside the detector.  Thus, for not too large $\tan\beta$, all the 
sleptons behave as stable charged particles from the point of view of 
the detector, and the signal consists of highly ionizing back-to-back 
tracks.  For moderately large $\tan\beta$, we expect a striking signal 
with $l^{+}l^{-} \tau^{\pm} \tau^{\mp} \tilde{\tau}^{\pm}
\tilde{\tau}^{\mp}$ in the final state, from slepton pair production 
and the subsequent three-body decays.

Another important aspect in the case where the bulk curvature is 
significant is that the gravitino is always the LSP. The gravitino 
mass is given by
\beq
\label{gravitinomass}
m_{3/2} = \frac{\tilde{F}}{\sqrt{3}M_{P}} = l_{4}^{1/2} l_{5}^{1/3}
\left(\frac{M_{5}}{k}\right) \left(\frac{M_c}{\sqrt{3}M_{P}}\right) 
M_{SUSY}~,
\eeq
where $\tilde{F} = F e^{-2k \LL}$ is the effective scale of SUSY 
breaking and we used Eq.~(\ref{susyF}) to estimate $F$, as well as 
Eq.~(\ref{4DPlanckscale}).  It is important that, unlike in the 
expression for the gaugino mass, Eq.~(\ref{gauginomass}), the 
denominator in Eq.~(\ref{gravitinomass}) is not redshifted by the warp 
factor, but rather is the four-dimensional Planck mass.  This can be 
most easily seen by writing the effective four-dimensional theory and 
canonically normalizing all the fields, including those responsible 
for supersymmetry breaking, which are localized on the IR brane.  We 
note that for $M_c \simeq 10^{6}~\rm{GeV}$, as in the third sample 
point in the table, Eq.~(\ref{gravitinomass}) gives $m_{3/2} \sim 
1~\rm{keV}$, and the gravitino becomes a good dark matter candidate.  
For higher compactification scales the gravitino is heavier and some 
means of gravitino dilution is necessary in order to avoid overclosure 
of the universe.  For sufficiently high compactification scales ($M_c 
\gsim 10^{14}~\rm{GeV}$, using the estimates given above), the 
gravitino becomes heavier than the other superpartners.  It is then 
necessary to ensure that the right-handed sleptons get additional 
contributions to make them heavier than the lightest neutralino.

\section{Conclusions}

Warped scenarios have several properties that make them very 
interesting for phenomenological applications.  Their most remarkable 
property is that they naturally contain vastly different scales.  In 
addition, when the gauge fields propagate in the bulk, the gauge 
couplings exhibit a logarithmic dependence on the fundamental 
parameters of the theory.  This makes possible the construction of 
weakly coupled grand unified models, which in turn can constrain the 
pattern of supersymmetry breaking.  Here we have concentrated on 
models in which the GUT symmetry is broken by a Higgs field localized 
on the UV brane and established that the physics localized on the 
infrared brane is GUT symmetric down to scales of order the infrared 
scale.  We found that the physical properties of these theories are 
more transparent when expressed in terms of the higher dimensonal 
propagators.  We computed the gauge boson and gaugino propagators in 
the case where the gauge symmetry is broken on the UV brane.  The 
present class of models provides a generalization of succesful 
supersymmetry breaking scenarios that solve the supersymmetric flavor 
problem by the sequestering mechanism. The warping effect, which is 
generically expected to be present in brane world scenarios, can have 
an important impact on the low energy phenomenology.

We showed that the ratios between the standard model gaugino masses 
depend on how close the compactification scale, $k\,e^{-k\LL}$, and 
the effective scale of supersymmetry breaking $M_{\rm{SUSY}}$ are.  
When the compactification scale is much higher than $M_{\rm{SUSY}}$ 
the physical gaugino masses are proportional to the corresponding 
gauge coupling squared.  However, this relation changes as the 
compactification scale approaches the weak scale.  In particular, when 
the SUSY breaking scale is much larger than the compactification scale, the 
gaugino masses are proportional to the gauge coupling.  In any case, 
the gauginos are not degenerate but exhibit a hierarchical pattern.

We have shown that all the SUSY breaking loop effects that are 
relevant in the observable sector are finite and calculable, being 
cutoff at the compactification scale.  When the compactification scale 
is much larger than the weak scale, the dominant contribution to the 
scalar masses is due to the RG running between these two scales.  
Generically, the right-handed stau is the NLSP, while the LSP is the 
gravitino.  When the compactification scale is close to the weak 
scale, the scalar masses are 1-loop suppressed compared to the gaugino 
masses.  In this case, other contributions due to hard SUSY breaking 
operators can easily be of the same size as the ones associated with 
the gaugino masses.  

\medskip

{\bf Acknowledgements} \\ 
We would like to thank Walter Goldberger, Markus Luty, Ann Nelson, 
Takemichi Okui, Maxim Perelstein and Tim Tait for discussions at 
various stages of this work.  We would also like to thank the Aspen 
Center for Physics for its hospitality.  Z.C. was supported 
by the Director, Office of Science, Office of High Energy and
Nuclear Physics, of the U.S. Department of Energy
under contract DE-ACO3-76SF00098 and 
by the N.S.F. under grant PHY-00-98840.  Fermilab is operated by 
Universities Research Association Inc.  under contract no.  
DE-AC02-76CH02000 with the DOE. 

Note added: While this paper was being completed 
we received \cite{CKKK} which considers related ideas.

\section{Appendix A}

In this appendix we give the gauge boson propagator when the gauge 
symmetry is broken by the VEV of a UV brane localized Higgs field.
For simplicity, we restrict here to the abelian case, but the results
generalize straightforwardly to the nonabelian case.  The effects
of the brane localized VEV will be treated exactly, since we are interested
in the case where $v \sim M_{GUT} \gg k\,e^{-k \LL}$.  We also restrict 
to the case where the brane localized gauge kinetic terms can be 
treated perturbatively, so that we can neglect them for the present 
calculation.  We assume that ${\cal{L}}_{UV}$ in Eq.~(\ref{action1}) 
contains terms like
\beq
{\cal{L}}_{UV} \supset 
- (D_\mu \phi)^\dagger D^\mu \phi - \lambda \left( |\phi|^2 - 
\frac{1}{2} v^2 \right)^2 ~,
\eeq
which induce a nonzero VEV, $\langle \phi \rangle = v/\sqrt{2}$.

It will be sufficient for our purposes to obtain the gauge boson 
propagator in $A_5 = 0$ gauge.  We will further choose 
four-dimensional unitary gauge, so that the localized gauge symmetry 
breaking appears simply as a localized mass for the gauge field.  In 
this case the gauge boson propagator satisfies
\beq
\frac{1}{g_5^2} \left\{ \eta^{\mu\nu} \partial^2 -\partial^\mu \partial^\nu +
\eta^{\mu\nu} \partial_y [e^{-2\sigma} \partial_y] -
g_5^2 v^{2}\,e^{-2\sigma} \eta^{\mu\nu} 2\delta(y) \right\}
G_{\nu\alpha}(X,X') = i \delta^\mu_\alpha \delta(X-X')~.
\eeq
Working in mixed position and 4-d (Euclidean) momentum 
space, the propagator can be written as
\beq
\label{Gmunu}
-i G_{\mu\nu}(y,y';p) = \left(\eta_{\mu\nu}-\frac{p_\mu p_\nu}{p^2}\right)
G_p(y,y') + \frac{p_\mu p_\nu}{p^2} G_0(y,y')~,
\eeq
where $G_p(y,y')$ satisfies
\beq
\label{Gpequation}
\frac{1}{g_5^2}\left[\partial_y \left(e^{-2\sigma}\partial_y G_p \right) -
p^2 G_p \right]-2\delta(y) v^2 e^{-2\sigma} G_p = \delta(y-y')~.
\eeq

Given that $A_{\mu}$ is even under the $Z_{2}$ orbifold parity, 
Eq.~(\ref{Gpequation}) imposes the following boundary conditions at $y 
= 0,
\LL$:
\beqa
\label{propbound0}
\left[ \partial_y G_p - m G_p \right]_{y = 0^{+}} &=& 0  \\
\label{propboundL}
\left. \partial_y G_p \right|_{y = \LL^{-}} &=& 0 ~,
\eeqa
The Higgs VEV only enters in the combination $m \equiv g_5^2 v^2$.
At $y = y'$, $G_p$ must be continuous and satisfy
\beq
\label{jump}
\left. e^{-2 \sigma} \partial_y G_{+} \right|_{y=y'+\epsilon} -
\left. e^{-2 \sigma} \partial_y G_{+} \right|_{y=y'-\epsilon} = g_{5}^{2}~,
\eeq
where $\epsilon \rightarrow 0^{+}$.  It is straightforward to find the 
solution to Eqs.~(\ref{Gpequation})-(\ref{jump}), which can be written 
as
\beq
\label{GaugeBosonPropagator}
G_p(y,y') = \frac{g_5^2}{k}e^{k(y+y')}
\frac{\left[A K_1\left(\frac{p}{k}e^{k y_<}\right)+
B I_1\left(\frac{p}{k}e^{k y_<}\right)\right]
\left[C K_1\left(\frac{p}{k}e^{k y_>} \right)+
D I_1\left(\frac{p}{k}e^{k y_>}\right) \right]}
{A D -B C}~,
\eeq
where, $K_{\alpha}$, $I_{\alpha}$ are modified Bessel functions,
$y_<$ ($y_>$) is the smallest (largest) of $y$, $y'$, and
\beqa
\label{coefficients}
A &=& I_0\left(\frac{p}{k}\right) - \frac{m}{p} I_1\left(\frac{p}{k}\right) \nonumber \\
B &=& K_0\left(\frac{p}{k}\right) + \frac{m}{p} K_1\left(\frac{p}{k}\right) \nonumber \\
C &=& I_0\left(\frac{p}{k}e^{k \LL}\right) \\
D &=& K_0\left(\frac{p}{k}e^{k \LL}\right)~. \nonumber
\eeqa

We also summarize here the approximate behavior of the propagator in 
different energy regimes, for arbitrary positions $y$, $y'$.  We 
assume that $m \equiv g_{5}^{2} v^{2} \lsim k$, but otherwise the 
following asymptotic forms are valid for any value of the symmetry 
breaking VEV. In particular, the propagator for an unbroken gauge 
field can be obtained by setting $m=0$.

i) $p \gg k \gsim m$
\beqa
\label{veryhighp}
G_p(y,y') &\sim& -\frac{g_5^2}{2p} \,
e^{\frac{1}{2}k(y+y')}
e^{-\frac{p}{k}( e^{k y_>} - e^{k y_<})}
\nonumber \\
& &\mbox{} \times \left[ 1 + e^{-\frac{2p}{k}(e^{k y_<}-1)} \right]
\left[ 1 + e^{-\frac{2p}{k}(e^{k \LL}-e^{k y_>})} \right]
\left[1 + {\cal{O}}\left(\frac{k}{p},\frac{m}{p}\right) \right]
\eeqa

ii) $k\,e^{-k y_<} \ll p \ll k$
\beqa
\label{highp}
G_p(y,y') &\sim& -\frac{g_5^2}{2p} \,
e^{\frac{1}{2}k(y+y')}
e^{-\frac{p}{k}( e^{k y_>} - e^{k y_<})}
\nonumber \\
& &\mbox{} \times \left[ 1 + e^{-\frac{2p}{k}(e^{k \LL}-e^{k y_>})} \right]
\left[1 + {\cal{O}}\left(\frac{p}{k},\frac{k}{p}e^{-ky_{<}}\right) \right]
\eeqa

iii) $k\,e^{-k y_>} \ll p \ll k\,e^{-k y_<}$
\beqa
\label{nonlocal}
G_p(y,y') &\sim& - \frac{g_5^2}{2} \,
e^{\frac{1}{2}k y_>} e^{-\frac{p}{k} e^{k y_>}}
\left[ 1 + e^{-\frac{2p}{k}(e^{k \LL}-e^{k y_>})} \right]  \nonumber \\
& & \times
\sqrt{\frac{\pi p}{2 k}} \; \frac{ 2 k + (e^{2 k y_<}-1) \, m }
{p^2 \ln(2k/p) + k \, m}
\left[1 + {\cal{O}}\left(\frac{p}{k}e^{ky_{<}}, \frac{k}{p}e^{-ky_{>}}\right) 
\right]
\eeqa

iv) $k\,e^{-k \LL} \ll p \ll k\,e^{-k y_>}$
\beq
\label{lowp}
G_p(y,y') \sim - \frac{g_5^2}{2} \;
\frac{2 k+(e^{2 k y_<}-1) \, m}{p^2 \ln(2k/p) + k \, m}
\left[1 + {\cal{O}}\left(\frac{k}{p}e^{-k \LL}, \frac{p}{k}e^{ky_{>}}\right) 
\right]
\eeq

v) $p \ll k\,e^{-k \LL}$
\beq
\label{verylowp}
G_p(y,y') \sim - \frac{g_5^2}{2 k} \;
\frac{2 k+(e^{2 k y_<}-1) \, m}{ \LL p^2 + \, m}
\left[1 + {\cal{O}}\left(\frac{p}{k}e^{k \LL}\right) 
\right]
\eeq

\section{Appendix B}

\newcommand{\Gll}{\ensuremath{G_{\GF\GF}}}
\newcommand{\Glp}{\ensuremath{G_{\GF\Psi}}}
\newcommand{\Gpl}{\ensuremath{G_{\Psi\GF}}}
\newcommand{\Gpp}{\ensuremath{G_{\Psi\Psi}}}

In this appendix we provide details of the derivation of the gaugino 
propagator in the case where the gauge symmetry is broken by the VEV 
of a Higgs field localized on the UV brane.  The left-handed (even) 
component of the gaugino field will then marry the Higgs fermion 
superpartner and obtain a localized Dirac mass.  We restrict again to 
the case where the gauge symmetry is abelian.  It is conventional to 
write the gaugino five-dimensional action in terms of a pair of 
symplectic Majorana spinors as in Eq.~(\ref{gauginoaction}) 
\cite{Mirabelli:1997aj}, which makes the $N=2$ $SU(2)$ R-symmetry of 
the original five-dimensional theory explicit.  However, for the 
purpose of deriving the propagator, it is simpler to write the bulk 
action in terms of a single five-dimensional Dirac spinor $\GF$.  Its 
left-handed components have the standard gauge interactions with the 
scalar and fermion components of the two chiral superfields $\Phi_+ = 
(\phi_+, \psi_+)$ and $\Phi_- = (\phi_-, \psi_-)$ involved in the 
breaking of the gauge symmetry.  Replacing these scalars by their VEVS 
gives a mass term that mixes the left-handed gaugino and one linear 
combination of the Weyl fermions $\psi_+$ and $\psi_-$.  For example, 
if $\langle \phi_+ \rangle = \langle \phi_- \rangle = v/2$ is 
real,\footnote{The normalization is chosen so that the effective VEV 
$\sqrt{\langle \phi_+ \rangle^2 + \langle \phi_- \rangle^2} = 
v/\sqrt{2}$ is normalized in the same way that was assumed in the 
derivation of the gauge boson propagator given in the previous 
appendix.} then $\psi = (\psi_- - \psi_+)/\sqrt{2}$.  We write this 
linear combination, $\psi$, and the orthogonal combination, $\chi$, in 
terms of a four-dimensional Dirac fermion $\Psi$.  The relation 
between these spinors and a two-component notation is
\beq
\label{spinors}
\begin{array}{cc}
\GF = \left( \begin{array}{c}
\lambda^{1}_{\alpha} \\ \bar{\lambda}_{2}^{\dot{\alpha}}
\end{array} \right) &
\hspace{1.5cm}
\Psi = \left( \begin{array}{c}
\chi_{\alpha} \\ \bar{\psi}^{\dot{\alpha}}
\end{array} \right)
\end{array}~.
\eeq
We neglect possible brane localized gaugino kinetic terms.  The
relevant part of the action reads then
\beqa
S &=& -\int_0^\LL d^4\!x\,dy \left\{ \sqrt{-G} \, \frac{i}{g_{5}^{2}} \left(
\overline{\GF}\,\Gamma^A {e_{\! A}}^{\! M} D_M \GF +
M \overline{\GF} \GF  \right) \right.
\nonumber \\
& & \left. \mbox{} + 2\delta(y) \, \sqrt{-G_{\rm{ind}}} \,
\left[ i \,\overline{\Psi}\,
\Gamma^a {e_{\! a}}^{\! \mu} D_\mu \Psi - \sqrt{2} i\,\frac{v}{\sqrt{2}}
\left( \overline{\GF} P_R \Psi -
\overline{\Psi} P_L \GF \right) \right] \right\}~,
\label{action}
\eeqa
where $D_M$ is a covariant derivative, with respect to both gauge and
general coordinate transformations, $G$ is the metric defined in
Eq.~(\ref{lineelement}), $G_{\rm{ind}}$ is the induced metric on the UV
brane, $e$ is the appropriate vielbein, and $M = c \sigma'$ is
an (odd) bulk mass term.  In the supersymmetric case $c = 1/2$.  We
use the following basis of $\Gamma$-matrices:
\beq
\begin{array}{cc}
\Gamma^\mu = \left( \begin{array}{cc}
0 & \sigma^\mu \\ \bar{\sigma}^\mu & 0
\end{array} \right) &
\hspace{1cm}
\Gamma^5 = \left( \begin{array}{cc}
1 & 0 \\ 0 & -1
\end{array} \right)
\end{array}~,
\eeq
where $\sigma^\mu = (1, \vec{\sigma})$, $\bar{\sigma}^\mu = (-1, 
\vec{\sigma})$, $\vec{\sigma}$ are the standard Pauli matrices, and 
$P_{L(R)} = \frac{1}{2} (1 \pm \Gamma^5)$.  The tree-level Green's 
function for the gaugino-higgsino system satisfies
\beq
\begin{array}{cc}
\left( \begin{array}{cc}
\frac{1}{g_{5}^{2}}  \cal{O}_{+}     & -i v P_{R} 2\delta(y)           \\
i v P_{L} 2\delta(y) & i \sla{\partial}\,2\delta(y)
\end{array}  \right)
\left( \begin{array}{cc}
\Gll  & \Glp           \\
\Gpl  & \Gpp
\end{array}  \right) =
\frac{i}{\sqrt{-G}}
\left( \begin{array}{cc}
1  & 0           \\
0  & 2\delta(y)
\end{array}  \right) \delta(X-X')
\end{array}~,
\eeq
where $\Gll = \langle\GF(X)\bar{\GF}(X')\rangle$,
$\Glp = \langle\GF(X)\bar{\Psi}(X')\rangle$, etc., and we defined
the operators
\beqa
\label{ops}
{\cal{O}}_{\pm} &\equiv& i \left( \Gamma^A {e_{\! A}}^{\! M}
D_M \pm M \right) \nonumber \\
 &=& i e^{\sigma} \sla{\partial} + i \Gamma^{5}
\partial_{y} - 2 i \sigma' \Gamma^{5} \pm i M~.
\eeqa
In 4-d momentum space, we then have to solve the system of first order
differential equations
\beqa
\label{eq1}
\frac{1}{g_{5}^{2}}{\cal{O}_{+}} \Gll - i v P_{R} 2\delta(y)
\Gpl &=& i e^{4\sigma} \delta(y-y') \\
\label{eq2}
\frac{1}{g_{5}^{2}}{\cal{O}_{+}} \Glp - i v P_{R} 2\delta(y)
\Gpp &=& 0\\
\label{eq3}
\left.\left( i v P_{L} \Gll + \sla{p} \Gpl \right)
\right|_{y=0} &=& 0\\
\label{eq4}
\left.\left( i v P_{L} \Glp + \sla{p} \Gpp \right)
\right|_{y=0} &=& 2i\delta(y')~,
\eeqa
where now $\Gll = \langle\GF(p;y)\bar{\GF}(-p;y')\rangle$, $\Glp =
\langle\GF(p;y)\bar{\Psi}(-p;y')\rangle$, etc.  We can use
Eqs.~(\ref{eq3}) and (\ref{eq4}) to find $\Gpl$ and $\Gpp$ at $y=0$,
and eliminate them from Eqs.~(\ref{eq1}) and (\ref{eq2}) to get
\beqa
\label{eqll}
{\cal{O}_{+}} \Gll - \frac{m}{\sla{p}} 2\delta(y) P_{L} \Gll &=&
i g_{5}^{2} e^{4\sigma} \delta(y-y') \\
\label{eqlp}
{\cal{O}_{+}} \Glp - \frac{m}{\sla{p}} 2\delta(y) P_{L} \Glp &=&
- \frac{i m}{\sla{p}} P_{L} 2\delta(y) \delta(y')~,
\eeqa
where we defined the mass parameter $m \equiv g_{5}^{2} v^{2}$ as we 
did in the previous appendix.

Since we are mainly interested in the $\Gll$ propagator, in the rest
of this appendix we concentrate on finding the solution $G \equiv
\Gll$ of Eq.~(\ref{eqll}), with ${\cal{O}_{+}}$ given in
Eq.~(\ref{ops}).  Due to the orbifold projection, the right- and
left-handed components of this propagator, defined by $G_{LL} \equiv
P_{L} G P_{R} = \langle \GF_{L} \overline{\GF_{L}} \rangle$, $G_{LR}
\equiv P_{L} G P_{L} = \langle \GF_{L} \overline{\GF_{R}} \rangle$,
etc., satisfy different boundary conditions and we need to study them
separately.  Applying $P_{L(R)}$ on the left of Eq.~(\ref{eqll}) and
$P_{R(L)}$ on the right, we get
\beq
\label{firstorder}
e^{\sigma} \sla{p}
\left( \begin{array}{c} G_{RL} \\ G_{LR} \end{array} \right)
+ (\pm i \partial_{y} \mp 2 i \sigma' + i c \sigma')
\left( \begin{array}{c} G_{LL} \\ G_{RR} \end{array} \right)
- \frac{m}{\sla{p}} 2\delta(y)
\left( \begin{array}{c} 0 \\ G_{LR} \end{array} \right) = 0~.
\eeq
We can also apply the operator ${\cal{O}_{-}}$ on the left of Eq.~(\ref{eqll})
and repeat the same projection operation. Using the identity
\beq
{\cal{O}_{-}}{\cal{O}_{+}} = e^{2\sigma} p^{2} - \partial_{y}^{2} +
i \sigma' e^{\sigma} \Gamma^{5} \sla{p} + 4 \sigma' \partial_{y}
- (4 - c^{2}) (\sigma')^{2} + (2 - c \Gamma^{5}) \sigma''~,
\eeq
we get a second set of equations
\beqa
\label{secondorder}
\left[e^{2\sigma} p^{2} - \partial_{y}^{2} + 4 \sigma' \partial_{y}
- (4 - c^{2}) (\sigma')^{2} + (2 \mp c) \sigma'' \right]
\left( \begin{array}{c} G_{LL} \\ G_{RR} \end{array} \right)
\hspace{3.2cm} \mbox
\nonumber \\
\pm i \sigma' e^{\sigma} \sla{p}
\left( \begin{array}{c} G_{RL} \\ G_{LR} \end{array} \right)
- m e^{\sigma} 2\delta(y)
\left( \begin{array}{c} G_{LL} \\ 0 \end{array} \right) +
\frac{m}{\sla{p}} i \left[ \partial_{y} - (2-c) \sigma'\right] 2\delta(y)
\left( \begin{array}{c} 0 \\ G_{LR} \end{array} \right)
\nonumber \\
= i g_{5}^{2} e^{5\sigma} \sla{p}
\left( \begin{array}{c} P_{R} \\ P_{L} \end{array} \right) \delta(y-y')~.
\eeqa
Equations~(\ref{firstorder}) and (\ref{secondorder}) couple $G_{LL}$
with $G_{RL}$ on the one hand, and $G_{RR}$ with $G_{LR}$ on the other.
We first solve for $G_{LL}$ and $G_{RL}$.  The upper line of
Eq.~(\ref{firstorder}) gives
\beq
\label{GRL}
G_{RL} = -i \frac{e^{-\sigma}}{\sla{p}} \left[ \partial_{y} - (2-c)
\sigma' \right] G_{LL}~,
\eeq
which can be used to eliminate $G_{RL}$ from the first line of
Eq.~(\ref{secondorder}):
\beqa
\left\{ \partial_{y}^{2} - e^{2\sigma} p^{2} - 5 \sigma' \partial_{y} +
[6 - c(c+1)] (\sigma')^{2} - (2 -c) \sigma'' - m 2\delta(y)  \right\}
G_{LL} \hspace{1.3cm} \mbox \nonumber \\
= - i g_{5}^{2} e^{5\sigma} P_{L} \sla{p} \delta(y-y')~.
\eeqa
This equation can be further simplified by writing $G_{LL} = i P_{L}
\sla{p} e^{2\sigma} G_{+}$:
\beq
\label{eqGeven}
\left\{ \partial_{y}^{2} - \sigma' \partial_{y} - e^{2\sigma} p^{2} -
c(c+1) (\sigma')^{2} + c \sigma'' - m 2\delta(y)  \right\}
G_{+} = - g_{5}^{2} e^{3\sigma} \delta(y-y')~.
\eeq
Using the fact that $G_{+}$ is even under the $Z_{2}$ orbifold, we now
can read the boundary conditions to be imposed on $G_{+}$.  For the
case of interest to us, where $c=1/2$, and using $\sigma'' =
2k[\delta(y)-\delta(y-\LL)]$, we have
\beqa
\label{boundaryeven1}
\left[ \partial_y G_{+} + \left(\frac{k}{2} - m \right) G_{+} 
\right]_{y = 0^{+}} &=& 0   \\
\label{boundaryeven2}
\left[ \partial_y G_{+} + \frac{k}{2} G_{+} \right]_{y = \LL^{-}} &=& 0~,
\eeqa
while at $y = y'$, $G_{+}$ must be continuous and satisfy
\beq
\label{boundaryeven3}
\left. e^{-3 \sigma} \partial_y G_{+} \right|_{y=y'+\epsilon} -
\left. e^{-3 \sigma} \partial_y G_{+} \right|_{y=y'-\epsilon} = - g_{5}^{2}~,
\eeq
where $\epsilon \rightarrow 0^{+}$.
The solution to Eq.~(\ref{eqGeven}), away from the boundaries and for
$y \neq y'$, is a linear combination of $e^{k \sigma/2} K_{1}(p \, e^{k
\sigma}/k)$ and $e^{k \sigma/2} I_{1}(p \, e^{k \sigma}/k)$, where
$K_{\alpha}$, $I_{\alpha}$ are modified Bessel functions.  It is
straightforward to impose the conditions Eqs.~(\ref{boundaryeven1}),
(\ref{boundaryeven2}) and (\ref{boundaryeven3}) to determine $G_{+}$.
Our final result for the $\langle \GF_{L} \overline{\GF_{L}} \rangle$
propagator is
\beq
\label{propagator}
G_{LL}(y,y') = - i e^{\frac{3}{2}\sigma} P_{L} \sla{p} G_{p}(y,y')~,
\eeq
where $G_{p}(y,y')$ is precisely the gauge boson propagator found in 
Eqs.~(\ref{GaugeBosonPropagator}) and (\ref{coefficients}).  
$G_{LR}(y,y')$ is then given by Eq.~(\ref{GRL}).

Now we turn to the solution for $G_{RR}$ and $G_{LR}$, which involves
some subtleties.  From the lower line of Eq.~(\ref{firstorder}),
\beq
\label{GLR}
\sla{p} e^{\sigma} G_{LR} = i \left[ \partial_{y} - (2+c)
\sigma' \right] G_{RR} + \frac{m}{\sla{p}} 2\delta(y) G_{LR}~,
\eeq
we notice that $G_{LR}$ should not have any $\delta$-function
singularities at $y=0$, or else the last term on the rhs of the
equation would be ill-defined.  It follows by integrating around a
small region about $y = 0$, and using the fact that $G_{RR}$
is odd, that
\beq
\label{jumpGRR}
G_{RR}(0^{+},y') - G_{RR}(0^{-},y') = 2 G_{RR}(0^{+},y') =
2i \frac{m}{\sla{p}} G_{LR} (0,y')~.
\eeq
If we now evaluate Eq.~(\ref{GLR}) at $y = 0$, using $\sigma'(0) = 0$
and the fact that (by virtue of Eq.~(\ref{jumpGRR})) the
$\delta$-functions on the rhs cancel, we find{\footnote{Note that
using Eq.~(\ref{jumpGRR}), we can write $\partial_{y} G_{RR}(y,y') =
F(y,y') + i (m/\sla{\!p}) G_{LR} 2\delta(y)$, where $F(y,y')$ is
regular at $y = 0$.  Then $F(0,y') = \partial_{y} G_{RR}(0^{+},y') =
\partial_{y} G_{RR}(0^{-},y')$.}}
\beq
\label{someequation}
\sla{p} G_{LR}(0,y') = i \partial_{y} G_{RR}(0^{+},y')~.
\eeq
Eliminating $G_{LR}(0,y')$ from Eqs.~(\ref{jumpGRR}) and
(\ref{someequation}), we get the following boundary condition for
$G_{RR}$ at $y=0^{+}$:{\footnote{It is important to distinguish
$G_{LR} (0,y')$ from $G_{LR} (0^{+},y') = G_{LR} (0^{-},y')$:
evaluating Eq.~(\ref{GLR}) at $y = 0^{+}$, and using
Eq.~(\ref{boundaryodd1}) to eliminate $G_{RL}(0^{+},y')$, followed by
Eq.~(\ref{jumpGRR}) to eliminate $\partial_{y} G_{RR}(0^{+},y')$, one
finds $G_{LR}(0^{+},y') = [1+(2+c)k m/p^{2}]G_{LR}(0,y')$.}}
\beq
\label{boundaryodd1}
G_{RR}(0^{+},y') = - \frac{m}{p^{2}} \partial_{y}G_{RR}(0^{+},y')~.
\eeq
We see that, as a consequence of the localized mass term, $G_{RR}$
does not vanish on the UV-brane.

Now we can restrict ourselves to $y \neq 0$, and find a second order
differential equation for $G_{RR}$ following the same steps that led
to Eq.~(\ref{eqGeven}) for $G_{LL}$: use Eq.~(\ref{GLR}) to
eliminate $G_{LR}$ from the second line of Eq.~(\ref{secondorder})
and define a new function $G_{-}$ by $G_{RR} = i P_{R} \sla{p}
e^{2\sigma} G_{-}$, which satisfies
\beq
\left\{ \partial_{y}^{2} - \sigma' \partial_{y} - e^{2\sigma} p^{2} -
c(c-1) (\sigma')^{2} + c \sigma''\right\}
G_{-} = - g_{5}^{2} e^{3\sigma} \delta(y-y')~.
\eeq

Integrating over a small region about $y = \LL$ and requiring that
$G_{-}$ be odd, leads to $G_{-}(L,y') = 0$.  Imposing this condition,
as well as Eq.~(\ref{boundaryodd1}), it is straightforward to find
the solution for $G_{-}$.  The $\langle \GF_{R} \overline{\GF_{R}}
\rangle$ propagator can finally be written as
\beqa
\label{oddpropagator}
G_{RR}(y,y') &=& i P_{R} \sla{p} \, \frac{g_5^2}{k} \,
\frac{e^{\frac{5}{2}k(y+y')}}{A_{-} D_{-} -B_{-} C_{-}}
\left[A_{-} K_0\left(\frac{p}{k}e^{k y_<}\right)-
B_{-} I_0\left(\frac{p}{k}e^{k y_<}\right)\right] \nonumber \\
& & \hspace{3.5cm} \times \left[C_{-} K_0\left(\frac{p}{k}e^{k y_>} \right)-
D_{-} I_0\left(\frac{p}{k}e^{k y_>}\right) \right]~,
\eeqa
where $y_<$ ($y_>$) is the smallest (largest) of $y$, $y'$ and
\beqa
\label{oddcoefficients}
A_{-} &=& I_0\left(\frac{p}{k}\right) - \frac{m/p}{1 + 5k m/(2p^{2})}
I_1\left(\frac{p}{k}\right) \nonumber \\
B_{-} &=& K_0\left(\frac{p}{k}\right) + \frac{m/p}{1 + 5k m/(2p^{2})}
K_1\left(\frac{p}{k}\right) \nonumber \\
C_{-} &=& I_0\left(\frac{p}{k}e^{k \LL}\right) \\
D_{-} &=& K_0\left(\frac{p}{k}e^{k \LL}\right)~. \nonumber
\eeqa
$G_{LR}(y,y')$ is then given by Eq.~(\ref{GLR}).

\section{Appendix C}

In this appendix we obtain an expression for the gaugino soft mass 
in the presence of gauge kinetic terms on the UV brane. The relevant
part of the action is
\beqa
S &=& -\int_0^\LL d^4\!x\,dy \left\{ \sqrt{-G} \, \frac{i}{g_{5}^{2}} \left(
\overline{\GF}\,\Gamma^A {e_{\! A}}^{\! M} D_M \GF +
M \overline{\GF} \GF  \right) \right. 
\nonumber \\
& & \left. \mbox{} + 2\delta \left( y \right ) \sqrt{-G_{\rm{ind}}}
\, \frac{i}{g_{UV}^{2}} 
\overline{\GF}\,\Gamma^{a} {e_{\! a}}^{\! \mu} D_{\mu} P_{L} \GF \right. 
\nonumber \\
& & \left. 
\mbox{} - 2\delta \left(y - \LL \right) \sqrt{-G_{\rm{ind}}} \,
\left[ \frac{1}{2} m_{IR}  \left( \GF^{T} C P_{L} \GF +
\overline{\GF} C P_{R} \overline{\GF}^{T} \right) \right] \right\}~,
\eeqa 
where $\GF$ is a Dirac spinor as in Eq.~(\ref{spinors}), $M = 
\sigma'/2$ and $C = - i \Gamma^{0} \Gamma^{2}$ is the charge 
conjugation operator.  The chirality projectors $P_{L}$, $P_{R}$ are 
inserted so that only the left handed components of $\GF$ receive 
terms on the branes.  The equations of motion are
\beqa 
\label{eqlambdaL}
i e^{\sigma} \sla{\partial} \GF_L 
- i \left( \partial_y - 2 \sigma' - M \right) \GF_R
&=& - i 2\delta(y) \frac{g_{5}^{2}}{g_{UV}^{2}} \sla{\partial} \GF_L 
- 2\delta(y-\LL) g_{5}^{2} m_{IR} C \Gamma^{0} \GF_{L}^{*}~,
\nonumber \\
\label{eqlambdaR}
i e^{\sigma} \sla{\partial} \GF_R 
+ i \left( \partial_y - 2 \sigma' + M \right) \GF_L
&=& 0~.
\eeqa
Working directly in 4-d momentum space, we can use the second equation 
to eliminate $\GF_R$ from the first one and get an equation for 
$\GF_{L}$.  We look for a solution of the form
\beq
\GF_L\left(x,y \right) = \eta_{L}(x)  A_+ \left( y \right)~,
\eeq
where $\eta_{L}(x) = e^{-i p x} \chi + e^{i p x}
(m/p^2) \sla{\!p} \,C \Gamma^0 \chi^*$ is a (4-component)
left-handed spinor satisfying the 4-dimensional Majorana equation
\beq
i \sla{\partial} \, \eta_{L} = - m C \Gamma^{0} \eta_{L}^* = m C 
\overline{\eta}_{L}^{T}~,
\eeq
and the function $A_{+}$ is real and even under the $Z_{2}$ orbifold.  
The resulting equation for $A_+$ is
\beqa
\label{eqAplus}
\left( \partial_y - 3 \sigma' - M \right) 
\left( \partial_y - 2 \sigma' + M \right) A_+
+ m^2 e^{2 \sigma} A_+ \nonumber \\
\mbox{} - 2\delta(y-\LL)
m g_{5}^{2} m_{IR} e^{\sigma} A_{+}
+ 2\delta(y) \frac{g_{5}^{2}}{g_{UV}^{2}} m^{2} A_+= 0~.
\eeqa

In the bulk, the solution has the general form
\beq
A_+ = e^{\frac{5}{2}\sigma} \left[ a J_1\left(\frac{m}{k}e^{\sigma}
\right) + b Y_1\left(\frac{m}{k}e^{\sigma} \right)
\right]~,
\eeq
where $J_1$ and $Y_1$ are Bessel functions and $a$ and $b$ are
determined by the boundary conditions.  The boundary conditions 
satisfied by $A_+$ can be read from Eq.~(\ref{eqAplus}): $A_+$ should 
be continuous at $y=0,L$ while its derivative satisfies
\beqa
\label{b0}
\partial_yA_+(0^{+}) &=& 
\left(\frac{3}{2}k - \frac{g_5^2}{g_{UV}^2} m^2 \right) A_+(0) \\
\label{bL}
\partial_yA_+(\LL^{-}) &=&
\frac{3}{2}k  A_+(L) - m \, m_{IR}\,g_5^2\,e^{k \LL}  A_+(\LL)~.
\eeqa
 
We consider first the limit $g_5 ^2 m_{IR} \ll 1$ and look for 
solutions with $m \ll k\,E^{-k \LL}$.  Using the expansions
\beqa
J_1(x) &\rightarrow& \frac{1}{2} x \nonumber \\
Y_1(x) &\rightarrow& \frac{1}{\pi} \left[ - \frac{2}{x} + 
x \ln x   - x \left( \ln 2 +
\frac{1}{2} - \gamma_{E} \right) \right]~,
\eeqa
we find that the lightest KK state has mass
\beq
m = g_4^2 m_{IR} e^{-k \LL}~,
\eeq
where $1/g_4^2 = L/g_5^2 + 1/g_{UV}^2$ determines the 4-d 
gauge coupling constant (characterizing the coupling of fermions to 
the zero-mode gauge bosons).  We see that the gaugino mass is 
proportional to the square of the four dimensional gauge coupling, as 
expected.

Next we consider the case where the SUSY breaking mass is large, $g_5 
^2 m_{IR} \gg \frac{k}{m} e^{- k \LL}$.  In this limit, we can 
replace Eq.~(\ref{bL}) by $A_+(L^-) = 0$, and expect the lowest 
eigenvalue to be of order $k\,e^{-k\LL}$.  It can be shown, however, 
that when $k \LL \gg 1$ the lightest state is parametrically smaller 
than $k\,e^{-k\LL}$ and it is possible to use the small argument 
approximation of the Bessel functions.  The result is
\beq
\label{largeSUSY}
m^2 = 2 k \frac{g_4^2}{g_5^2} e^{-2 k \LL} \left[ 1+
\frac{3}{4} \frac{g_4^2}{k g_5^2} +
{\cal{O}} \left( \frac{g_4^2}{k g_5^2}\right)^2 \right]~.
\eeq
We see that the gaugino mass is now proportional to the four 
dimensional gauge coupling instead of the square of the gauge coupling 
as in the previous case.


\medskip

\begin{thebibliography}{99}

\bibitem{Randall:1998uk}
L.~Randall and R.~Sundrum,
Nucl.\ Phys.\ B {\bf 557}, 79 (1999)
[arXiv:hep-th/9810155].

\bibitem{RS1}
L.~Randall and R.~Sundrum,
Phys.\ Rev.\ Lett.\  {\bf 83}, 3370 (1999)
[arXiv:hep-ph/9905221].

\bibitem{Giudice:1998xp}
G.~F.~Giudice, M.~A.~Luty, H.~Murayama and R.~Rattazzi,
JHEP {\bf 9812}, 027 (1998)
[arXiv:hep-ph/9810442].

\bibitem{Kaplan:1999ac}
D.~E.~Kaplan, G.~D.~Kribs and M.~Schmaltz,
Phys.\ Rev.\ D {\bf 62}, 035010 (2000)
[arXiv:hep-ph/9911293].

\bibitem{Chacko:1999mi}
Z.~Chacko, M.~A.~Luty, A.~E.~Nelson and E.~Pont\'{o}n,
JHEP {\bf 0001}, 003 (2000)
[arXiv:hep-ph/9911323].

\bibitem{Chacko:2000fn}
Z.~Chacko and M.~A.~Luty,
JHEP {\bf 0105}, 067 (2001)
[arXiv:hep-ph/0008103].

\bibitem{Kaplan:2000jz}
D.~E.~Kaplan and G.~D.~Kribs,
JHEP {\bf 0009}, 048 (2000)
[arXiv:hep-ph/0009195].

\bibitem{Anisimov:2001zz}
A.~Anisimov, M.~Dine, M.~Graesser and S.~Thomas,
Phys.\ Rev.\ D {\bf 65}, 105011 (2002)
[arXiv:hep-th/0111235].

\bibitem{Anisimov:2002az}
A.~Anisimov, M.~Dine, M.~Graesser and S.~Thomas,
arXiv:hep-th/0201256.

\bibitem{Goldberger:1999wh}
W.~D.~Goldberger and M.~B.~Wise,
Phys.\ Rev.\ D {\bf 60}, 107505 (1999)
[arXiv:hep-ph/9907218].
 
\bibitem{hewettpomarol}
H.~Davoudiasl, J.~L.~Hewett and T.~G.~Rizzo,
Phys.\ Lett.\ B {\bf 473}, 43 (2000)
[arXiv:hep-ph/9911262];
A.~Pomarol,
Phys.\ Lett.\ B {\bf 486}, 153 (2000)
[arXiv:hep-ph/9911294].

\bibitem{Pomarol:2000hp}
A.~Pomarol,
Phys.\ Rev.\ Lett.\  {\bf 85}, 4004 (2000)
[arXiv:hep-ph/0005293].

\bibitem{Randall:2001gb}
L.~Randall and M.~D.~Schwartz,
JHEP {\bf 0111}, 003 (2001)
[arXiv:hep-th/0108114];
L.~Randall and M.~D.~Schwartz,
Phys.\ Rev.\ Lett.\  {\bf 88}, 081801 (2002)
[arXiv:hep-th/0108115].
 
\bibitem{Contino:2002kc}
R.~Contino, P.~Creminelli and E.~Trincherini,
JHEP {\bf 0210}, 029 (2002)
[arXiv:hep-th/0208002].
 
\bibitem{Choi:2002zi}
K.~w.~Choi, H.~D.~Kim and I.~W.~Kim,
arXiv:hep-ph/0207013.
 
\bibitem{Goldberger:2002cz}
W.~D.~Goldberger and I.~Z.~Rothstein,
Phys.\ Rev.\ Lett.\  {\bf 89}, 131601 (2002)
[arXiv:hep-th/0204160];
W.~D.~Goldberger and I.~Z.~Rothstein,
arXiv:hep-th/0208060.

\bibitem{Agashe:2002bx}
K.~Agashe, A.~Delgado and R.~Sundrum,
Nucl.\ Phys.\ B {\bf 643}, 172 (2002)
[arXiv:hep-ph/0206099].
 
\bibitem{Falkowski:2002cm}
A.~Falkowski and H.~D.~Kim,
JHEP {\bf 0208}, 052 (2002)
[arXiv:hep-ph/0208058];
\bibitem{Randall:2002qr}
L.~Randall, Y.~Shadmi and N.~Weiner,
arXiv:hep-th/0208120.
 
\bibitem{Choi:2002ps}
K.~w.~Choi and I.~W.~Kim,
arXiv:hep-th/0208071.
 
\bibitem{Agashe:2002jx}
K.~Agashe and A.~Delgado,
arXiv:hep-th/0209212.

\bibitem{Agashe:2002pr}
K.~Agashe, A.~Delgado and R.~Sundrum,
arXiv:hep-ph/0212028.

\bibitem{GP1}
T.~Gherghetta and A.~Pomarol,
Nucl.\ Phys.\ B {\bf 586}, 141 (2000)
[arXiv:hep-ph/0003129].

\bibitem{GP2}
T.~Gherghetta and A.~Pomarol,
Nucl.\ Phys.\ B {\bf 602}, 3 (2001)
[arXiv:hep-ph/0012378].

\bibitem{Marti:2001iw}
D.~Marti and A.~Pomarol,
Phys.\ Rev.\ D {\bf 64}, 105025 (2001)
[arXiv:hep-th/0106256].

\bibitem{Goldberger:2002pc}
W.~D.~Goldberger, Y.~Nomura and D.~R.~Smith,
arXiv:hep-ph/0209158.

\bibitem{Arkani-Hamed:2001ca}
N.~Arkani-Hamed, A.~G.~Cohen and H.~Georgi,
Phys.\ Rev.\ Lett.\  {\bf 86}, 4757 (2001)
[arXiv:hep-th/0104005].

\bibitem{Hill:2000mu}
C.~T.~Hill, S.~Pokorski and J.~Wang,
Phys.\ Rev.\ D {\bf 64}, 105005 (2001)
[arXiv:hep-th/0104035].

\bibitem{Carena:2002me}
M.~Carena, T.~M.~Tait and C.~E.~Wagner,
arXiv:hep-ph/0207056.

\bibitem{Carena:2002dz}
M.~Carena, E.~Pont\'{o}n, T.~Tait and C.~E.~Wagner,
arXiv:hep-ph/0212307.

\bibitem{Manohar:1996cq}
A.~V.~Manohar,
arXiv:hep-ph/9606222.

\bibitem{WessBagger}
J.~Wess and J.~Bagger, ``Supersymmetry and Supergravity'', Second 
Edition, Princeton University Press (1992).

\bibitem{Chacko:1999hg}
Z.~Chacko, M.~A.~Luty and E.~Pont\'{o}n,
JHEP {\bf 0007}, 036 (2000)
[arXiv:hep-ph/9909248].

\bibitem{Arkani-Hamed:2001mi}
N.~Arkani-Hamed, L.~J.~Hall, Y.~Nomura, D.~R.~Smith and N.~Weiner,
Nucl.\ Phys.\ B {\bf 605}, 81 (2001)
[arXiv:hep-ph/0102090].

\bibitem{Giudice:1998bp}
G.~F.~Giudice and R.~Rattazzi,
Phys.\ Rept.\  {\bf 322}, 419 (1999)
[arXiv:hep-ph/9801271].

\bibitem{Ambrosanio:1997bq}
S.~Ambrosanio, G.~D.~Kribs and S.~P.~Martin,
Nucl.\ Phys.\ B {\bf 516}, 55 (1998)
[arXiv:hep-ph/9710217].

\bibitem{Mirabelli:1997aj}
E.~A.~Mirabelli and M.~E.~Peskin,
Phys.\ Rev.\ D {\bf 58}, 065002 (1998)
[arXiv:hep-th/9712214].

\bibitem{CKKK}
K.~Choi, D.~Y.~Kim, I.~W.~Kim and T.~Kobayashi,
[arXiv:hep-ph/0301131].

\end{thebibliography}
\end{document}